\documentclass[aps,prd,preprint,floatfix,superscriptaddress,nofootinbib,longbibliography,a4paper]{revtex4-1}
\pdfoutput=1
\usepackage{color}
\usepackage{graphicx}
\usepackage{textcomp}
\usepackage{subfigure}
\usepackage{feynmp}
\usepackage{amsmath,amssymb,array}
\allowdisplaybreaks
\usepackage{enumerate}
\usepackage{enumitem}
\usepackage{hhline}
\usepackage{hyperref}
\usepackage{orcidlink}
\usepackage{comment}
\usepackage{siunitx}
\usepackage{multirow}
\hypersetup{
    citecolor= red,
    colorlinks=true,
    linkcolor=blue,
    filecolor=magenta,      
    urlcolor=blue,}

\usepackage{epsfig}
\usepackage{xfrac}
\usepackage{slashed}
\usepackage{pifont}

\newcommand{\beqa}{\begin{eqnarray}}
\newcommand{\eeqa}{\end{eqnarray}}
\newcommand{\be}{\begin{equation}}
\newcommand{\ee}{\end{equation}}
\newcommand{\ba}{\begin{array}} 
\newcommand{\ea}{\end{array}}

\newcommand{\nl}{\nonumber \\}
\newcommand{\cc}{\, {\mathcal{C}}^{-1}\,}
\newcommand{\hc}{\;+\; \mathrm{H.c.}}
\newcommand{\hcn}{& + & \mathrm{H.c.}}

\newcommand{\ft}{\mathbf{10}}
\newcommand{\ff}{\mathbf{\overline{5}}}

\newcommand{\hh}{\mathrm{H}}
\newcommand{\eq}{\,&=&\,}
\newcommand{\ad}{\,&+&\,}
\newcommand{\mi}{\,&-&\,}

\begin{document} 
\vspace*{0.5cm}
\title{Revisiting $SU(5)$ Yukawa Sectors Through Quantum Corrections}
\bigskip

\author{Saurabh K. Shukla \orcidlink{0000-0001-5344-9889}}
\email{saurabhks@nankai.edu.cn}
\affiliation{School of Physics, Nankai University, Tianjin 300071, China\vspace*{1cm}}

\begin{abstract}

This article revisits the validity of tree-level statements regarding the Yukawa sector of various minimal-renormalisable \(SU(5)\) frameworks at the loop level. It is well-known that an \(SU(5)\) model with only the \(45_{\mathrm{H}}\) dimensional irreducible representation~(irrep) contributing to the Yukawa sector is highly incompatible in yielding the low-energy observables. However, this study shows that when one-loop corrections from heavy degrees of freedom are included in the various Yukawa vertices, the model can reproduce the charged fermion mass spectrum and mixing angles within the assumed experimental uncertainty. The fitted Yukawa couplings remain within the perturbative range. The scenario also necessitates mass splitting among various scalars of $45_{\mathrm{H}}$ dimensional irrep, with some of the scalars' mass deviating significantly from the matching scale \((M_{GUT})\), collectively providing substantial threshold corrections. The effect of the lightest scalar on the RGE evolution of the SM parameters has also been considered. It is shown that this scenario reproduces the SM spectrum at low energy. As an extension, the minimal \(SU(5)\) model with only the \(45_{\mathrm{H}}\) irrep is augmented with the \(15_{\mathrm{H}}\)-dimensional irrep, which also successfully reproduces the observed charged and neutral fermion mass spectra. Finally, the study considers an alternative \(SU(5)\) model incorporating both \(5_{\mathrm{H}}\) and \(15_{\mathrm{H}}\) irreps, which also yields the desired fermion mass spectra and mixing angles. This work demonstrates the viability of a minimal \(SU(5)\) Yukawa sector in different setups when quantum corrections are considered.

\end{abstract}

\maketitle

\section{Prelude}\label{sec:bg}
Grand Unified Theories (GUTs) are well-motivated frameworks in dealing with some of the inconsistencies of the Standard Model (SM)~\cite{Georgi:1974sy,Fritzsch:1974nn}. Once the quarks and leptons are unified in a common irreducible representation~(irrep) and GUTs offer a plethora of irreps comprising of distinct scalar fields, in addition to SM Higgs~\cite{Gell-Mann:1979vob,Langacker:1980js,Nath:2006ut}.\\ 

A grand unification model with only a single irrep in the Yukawa sector often leads to inconsistent tree-level Yukawa relations.  For example, in an \(SU(5)\) model with a single scalar irrep \(5_{\hh}\) contributing to the Yukawa sector, the relation \(Y_d = Y_e^T\) arises, which is observationally inconsistent. To resolve this, multiple scalar irreps are typically introduced, modifying the tree-level Yukawa relations and making them phenomenologically viable~\cite{Georgi:1979df,Barbieri:1981yw,Giveon:1991zm,Dorsner:2006dj,FileviezPerez:2007bcw,Goto:2023qch}. However, this approach introduces the hierarchy problem and numerous tunable parameters in the scenario, which are one of the most pronounced problems of the GUTs. Alternative solutions involve incorporating higher-dimensional operators~\cite{Ellis:1979fg,Berezinsky:1983va,Altarelli:2000fu,Emmanuel-Costa:2003szk,Dorsner:2006hw,Antusch:2021yqe,Antusch:2022afk} or vector-like fermions~\cite{Hempfling:1993kv,Shafi:1999rm,Barr:2003zx,Oshimo:2009ia,Babu:2012pb,Dorsner:2014wva,FileviezPerez:2018dyf,Antusch:2023mxx,Antusch:2023kli,Antusch:2023mqe}, both of which alter the tree-level matching conditions. The inconsistency in Yukawa relations for minimal\footnote{Minimality here refers to having only a single irrep consisting of SM Higgs contributing to the Yukawa sector.} GUT models are not exclusive to \(SU(5)\) but also applies to \(SO(10)\) GUTs. However, our findings suggest that a minimal \(SU(5)\) model with only \(5_{\hh}\) in the Yukawa sector, extended with \(SU(5)\)-singlet fermion(s), can yield a realistic fermion mass spectrum and mixing angles when quantum corrections are included~\cite{Patel:2023gwt}.  \\

This study, on the same line of~\cite{Patel:2023gwt}, aims to comprehensively investigate various minimal \(SU(5)\) frameworks with different irreps in the Yukawa sector, assessing their potential to yield a viable fermion mass spectrum when quantum corrections are included. Such an investigation is essential for constructing a minimal and calculable GUT model. In \(SU(5)\), only two irreps—\(5_{\hh}\) and \(45_{\hh}\)—can couple to the matter multiplets at the renormalisable level, with each containing a submultiplet transforming like the SM Higgs. Minimal versions that include only one of these irreps, i.e. $5_{\hh}$ and $45_{\hh}$, in the Yukawa sector are known to produce inconsistent tree-level Yukawa relations, as previously discussed. The present work revisits the validity of this tree-level statement at the loop level. \\

The central focus of this study is to evaluate the viability of Yukawa sectors where only a single irrep containing the SM Higgs contributes to the Yukawa sector. It examines various renormalisable \(SU(5)\) frameworks— with \(45_{\hh}\), \(45_{\hh}-15_{\hh}\), and \(5_{\hh}-15_{\hh}\) irreps—regarding their potential to reproduce the observed fermion mass spectra and mixing angles when loop corrections to tree-level Yukawa vertices are incorporated. The quantum corrections arise from the heavy degrees of freedom inherent to the models, which are various scalar fields arising from the same scalar irrep and the heavy gauge boson. To provide sufficient threshold corrections at the matching scale, a mass splitting among the scalar masses is necessary, which extends beyond the Extended Survival Hypothesis~(ESH)~\cite{delAguila:1980qag,Mohapatra:1982aq,Dimopoulos:1984ha}. The ESH suggests that only the scalars necessary for breaking a symmetry at a given energy scale are present at that scale and can remain light. The consequence of $45_{\hh}$ dimensional irrep in contributing to fermion masses~\cite{Babu:1984vx,FileviezPerez:2007bcw,FileviezPerez:2016sal}, gauge coupling unification~\cite{Dorsner:2006dj,Haba:2024lox,Fang:2024mfn} and flavour anomalies~\cite{Becirevic:2018afm,Goto:2023qch} are already present in the literature. Additionally, $15_{\hh}$ dimensional irrep is long known to contribute to neutrino mass~\cite{Dorsner:2007fy,Dorsner:2017wwn} via type-II seesaw mechanism~\cite{Schechter:1980gr,Mohapatra:1980yp}. \\

This article is organised as follows: Section~(\ref{sec:allmodels}) begins with a discussion on one loop matching condition of the Yukawa sector, followed by subsections~(\ref{ssec:modelI}, \ref{ssec:modelII}, and \ref{ssec:modelIII}), which examine different \(SU(5)\) models in various frameworks: \(45_{\hh}\), \(45_{\hh}-15_{\hh}\), and \(5_{\hh}-15_{\hh}\), respectively, and calculate the one-loop corrections to various Yukawa vertices. Section~(\ref{sec:analysis}) presents the numerical analysis of the frameworks discussed in Section~(\ref{sec:allmodels}), evaluating the effectiveness of the one-loop-corrected Yukawas in reproducing the largest Yukawas in the charged fermion sectors. Section~(\ref{sec:solutions}) provides the benchmark values of the fitted parameters for the various \(SU(5)\) models with different Yukawa sectors considered in the study. The section~(\ref{sec:summary}) summarised the carried out study and appendix~(\ref{app:LF}) is included to support the contents of the main text.

\section{$SU(5)$ Yukawa Relations at one loop}
\label{sec:allmodels}

Consider a generic $SU(5)$ model with three generations of $\ff$ and $\ft$ irreps consisting of SM quarks and leptons. The SM fermionic embedding of the $SU(5)$ matter multiplets, i.e. $\ff$ and $\ft$, is shown below~\cite{Nath:2006ut}.
\beqa{\label{eq:ff&ft}}
\ff_{a}\eq \varepsilon_{ab}\,l^b,\hspace{0.5cm}\ff_{\alpha}\;=\;d^C_{\alpha}\,,\nl
\ft^{a\alpha} \eq \frac{1}{\sqrt{2}}\, q^{a\,\alpha},\hspace{0.5cm}\ft^{\alpha\beta}\;=\;\frac{1}{\sqrt{2}}\,\varepsilon^{\alpha\beta\gamma}\,u^C_{\gamma},\hspace{0.5cm}\ft^{ab}\;=\;\,\frac{1}{\sqrt{2}}\varepsilon^{ab}\,e^C\,,
\eeqa
where, Greek letters $\big(1\leq\,\,\alpha,\, \beta,\,\gamma...\,\,\leq 3)$ denotes $SU(3)_{\rm C}$ indices while $SU(2)_{\rm{L}}$ labels are depicted by the lowercase Latin alphabets $(4\leq a,b,c... \leq 5 )$. The convention of two-indexed Levi-Civita tensor is as follows: $\varepsilon_{45}\,=1\,=\,\varepsilon^{54}\,=\,-\varepsilon_{54}\,=\,-\varepsilon^{45}$. Further, three indexed Levi-Civita follows the convention where $\varepsilon_{123}\,=\,1$ and for other cyclic permutations. The generation labels are designated by upper case Latin alphabets, i.e. $\big( 1 \,\leq\, A,\,B,\,C,\,...\leq 3\big)$.\\

\begin{figure}[t!]
    \centering
    \includegraphics[width=0.75\linewidth]{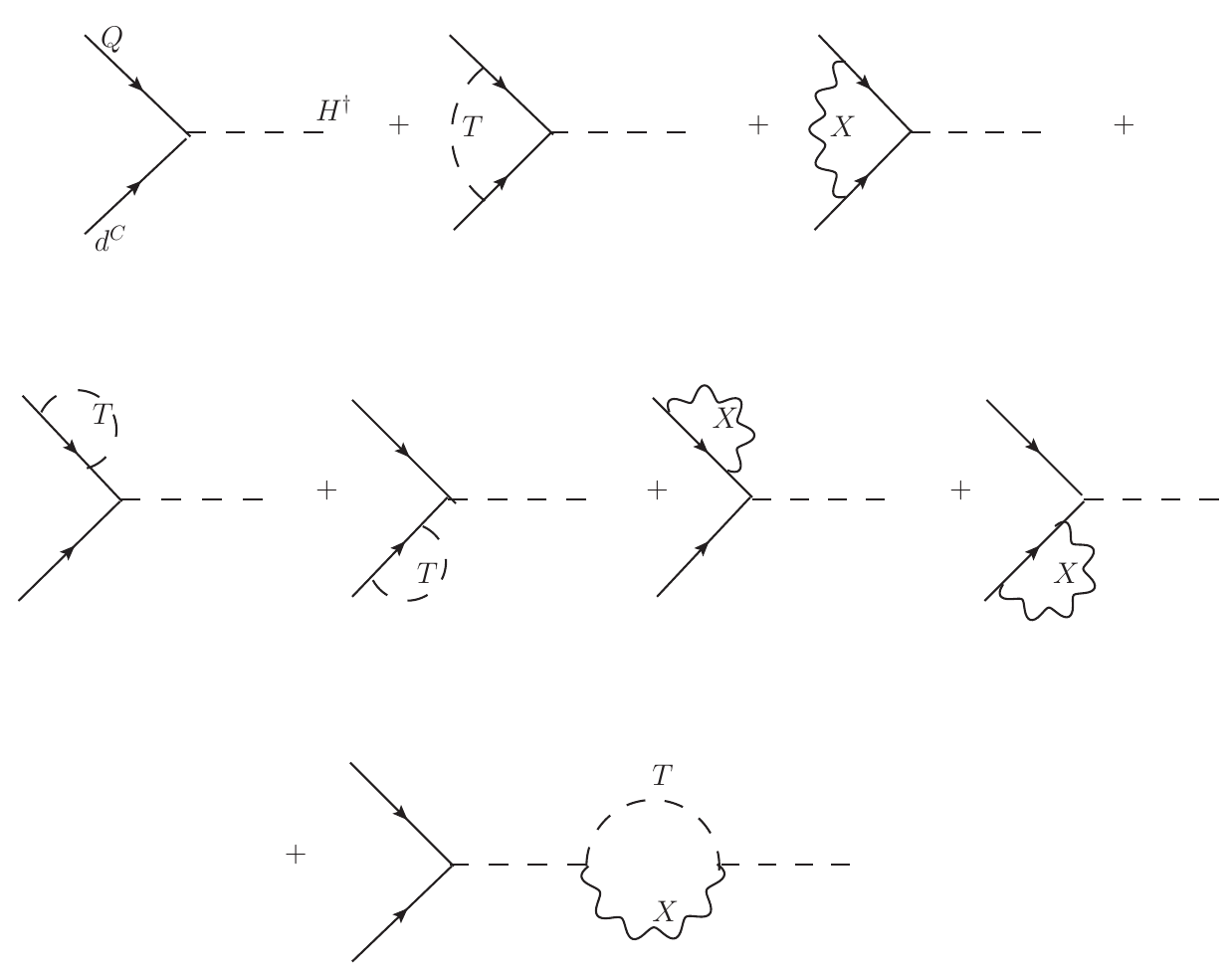}
    \caption{One-loop correction diagrams for \(Y_d\) are shown, with analogous diagrams also applying to \(Y_e\) and \(Y_u\). Here, \(T\) denotes a scalar field transforming in an irreducible representation that participates in the Yukawa interactions, while \(X\) denotes the gauge boson. The first line of diagrams represents the vertex corrections, the second line represents the wave-function corrections to the fermion legs arising from different scalar fields and heavy gauge bosons, and the last line represents the wave-function correction to the Higgs leg due to scalar fields and gauge bosons.}
    \label{fig:oneloopfigs}
\end{figure}

This section discusses different versions of \(SU(5)\) models—namely, the \(45_H\), \(45_H\text{-}15_H\), and \(5_H\text{-}15_H\) models—and evaluates the one-loop corrections to the tree-level Yukawa vertices. It is to be noted that the \(50_H\)-dimensional \(SU(5)\) irreducible representation also couples to the \(SU(5)\) fermion multiplets at the renormalisable level; however, it does not contain the SM Higgs doublet~\cite{Slansky:1981yr}.  Following Refs.~\cite{Weinberg:1980wa,Hall:1980kf,PhysRevD.28.194,Kane:1993td,Hempfling:1993kv,Wright:1994qb}, the one-loop matching condition for the Yukawa couplings at a given renormalisation scale \(\mu\) was derived in Ref.~\cite{Patel:2023gwt} and is given by
\beqa{\label{eq:matchcond}}
Y_f\big(\mu\big) \eq Y_f^0\left(1-\frac{K_H\big(\mu\big)}{2}\right) + \delta Y_f\big(\mu\big) - \frac{1}{2}\left(K_f^T\,\big(\mu\big)\,Y_f^0 + Y_f^0\,K_f^C\,\big(\mu\big)\right),
\eeqa
where \(Y_f^0\) denotes the tree-level Yukawa coupling of the fermions \(f\,\supset\,\{q,\,u^C,\,d^C,\,l,\,e^C\}\) to the SM Higgs. The one-loop corrected Yukawa coupling at the scale \(\mu\), namely \(Y_f\big(\mu\big)\), is fully determined in terms of the finite part of the vertex correction, \(\delta Y_f\), and the wave-function renormalisation factors \(K_H\) and \(K_f\). The relevant one-loop diagrams for the vertex $Y_d$ are shown in Fig.~\ref{fig:oneloopfigs}, where the first row corresponds to vertex corrections, the second row to wave-function corrections to the fermion legs induced by scalar fields and heavy gauge bosons, and the third row to the wave-function correction to the Higgs leg.

\subsection{ Model I : $SU(5)$ with only $45_{\hh}$}
\label{ssec:modelI}
The scalar sector of this $SU(5)$ model consists only of $45_{\hh}$-dimensional irrep. The various scalar fields residing in $45_{\hh}$ are listed in Tab.~(\ref{tab:scalars45H}). The decomposition of $45_{\hh}$-plet into various irreps with canonically normalised kinetic terms can be inferred from~\cite{Patel:2022wya} and is reproduced below for convenience:

\beqa \label{eq:45dec}
45_{\hh\,\gamma}^{\alpha \beta} &\equiv& S_{\gamma}^{\alpha \beta} + \frac{1}{2 \sqrt{2}} \left( \delta_\gamma^{\alpha} T^\beta - \delta_\gamma^\beta T^\alpha \right)\,,~~~~45_{\hh\,a}^{\alpha \beta}  \equiv   \Omega_a^{\alpha \beta}\,\,, \nonumber \\
45_{\hh\,\beta}^{\alpha a} &\equiv& \frac{1}{\sqrt{2}} O_\beta^{\alpha a} + \frac{1}{2\sqrt{6}} \delta_\beta^\alpha H^a\,,~~~~45_{\hh\,\beta}^{ab} \equiv \frac{1}{\sqrt{2}} \varepsilon^{ab} {\cal T}^{\dagger}_\beta\,,\nonumber \\
45_{\hh\,a}^{b \alpha} & \equiv & \frac{1}{\sqrt{2}} \mathbb{T}_a^{b \alpha} - \frac{1}{2 \sqrt{2}} \delta_a^b T^\alpha\,,~~~~ 45_{\hh\,c}^{ab} \equiv -\frac{\sqrt{3}}{2\sqrt{2}} \left( \delta_c^a H^b - \delta_c^b H^a\right)\,. \eeqa

\begin{table}[t]
    \centering
    \begin{tabular}{cc}
    \hline\hline
        ~~~~~~~Notation~~~~~~~ & ~~~~~~~SM Charge~~~~~~~ \\\hline
         $H$& $\big(1,2,\frac{1}{2}\big)$ \\
        $T$ & $\big(3,1,-\frac{1}{3}\big)$\\
        ${\cal T}$ & $\big(\overline{3},1,\frac{4}{3}\big)$ \\
        $\Omega$ & $\big(\overline{3},2,-\frac{7}{6}\big)$ \\
        $\mathbb{T}$ & $\big(3,3,-\frac{1}{3}\big)$ \\
        $S$ & $\big(\overline{6},1,-\frac{1}{3}\big)$ \\
        $O$ & $\big(8,2,\frac{1}{2}\big)$ \\
\hline\hline
    \end{tabular}
    \caption{Scalar multiplets residing inside $45_{\hh}$ dimensional irrep.}
    \label{tab:scalars45H}
\end{table}

The interaction of $\ff$ and $\ft$ with $45_\hh$-dimensional field at renormalisable level is parametrised as follows:
\beqa{\label{eq:SU5Tree}}
-\mathsf{L}^{45_{\hh}}_{\mathrm{Y}} \eq \frac{1}{2}\,\big(Y_{1}\big)_{AB}\,\ft_{A}^T\,\cc\ft_B\,45_{\hh} \,+\, \sqrt{2}\,\big(Y_{2}\big)_{AB}\,\ft_{A}^T\,\cc\,\ff_B\,45_{\hh}^{\dagger}\, \hc\,, 
\eeqa
where $Y_1$ is antisymmetric in flavour indices. Using the dictionary provided in Eq.~(\ref{eq:ff&ft}) together with the canonically normalised decomposition of $45_{\hh}$ given in Eq.~(\ref{eq:45dec}), the vertices involving SM-Higgs $(H)$ and fermions can be evaluated as follows:
\beqa {\label{45H-SM}}
-\mathsf{L}^{45_{\hh}}_{\mathrm{Y}} &\supset&  - Y_{1\,AB}\,\frac{2}{\sqrt{6}}\,\varepsilon_{ab}\,q^{a\alpha\,T}_A\,\cc\,u^C_{\alpha\,B}\,H^b \nl \mi Y_{2\,AB}\,\Big(\frac{1}{\sqrt{6}}\,q^{a\alpha\,T}_A\,\cc\,d^C_{\alpha\,B}\,H^{\dagger}_a - \sqrt{\frac{3}{2}}\,e^{C\,T}_A\,\cc\,l_B^a\,H^{\dagger}_a \Big)
 \hc\eeqa

The interaction terms written in Eq.~(\ref{45H-SM}) lead to the following tree-level Yukawa relations valid at the GUT scale:
\beqa{\label{eq:tree-yuk}}
\big(Y_{u}\big)_{AB} \eq \frac{2}{\sqrt{6}}\,\big(Y_{1}\big)_{AB}\,,\nl
\big(Y_{d}\big)_{AB} \eq \frac{1}{\sqrt{6}}\,\big(Y_{2}\big)_{AB}\,,\nl
\big(Y_{e}\big)_{AB} \eq -\sqrt{\frac{3}{2}}\,\big(Y_{2}^T\big)_{AB}\,.
\eeqa

The tree-level Yukawa relations mentioned in Eq.~(\ref{eq:tree-yuk}) are inconsistent due to the following reasons:
\begin{enumerate}
    \item Inconsistent Yukawa relations in down type and charged lepton sector. {\label{pt:aa}}\\
The Yukawa relations mentioned in Eq.~(\ref{eq:tree-yuk}) lead to $3\,Y_d\,=-\,Y_e^T$ which result in $\frac{y_d}{y_e}\,=\,\frac{y_c}{y_{\mu}}\,=\,\frac{y_b}{y_{\tau}}\,=\,\frac{1}{3}$. However, at the traditional GUT scale~$(M_{\rm{GUT}} \sim 10^{16} \, \text{GeV})$, the RGE extrapolated SM values requires $\frac{y_d}{y_e} = 2$, $\frac{y_c}{y_{\mu}} = \frac{1}{5}$, and $\frac{y_b}{y_{\tau}} = \frac{2}{3}$.
\item Inaccurate spectrum in the up-quark sector.\label{pt:bb}\\
The antisymmetric nature of $Y_1$ in three dimensions makes its determinant zero, implying at least one of the three eigenvalues is zero. Moreover, this scenario also predicts two degenerate eigenvalues at the tree-level.
    
\end{enumerate}

This subsection aims to investigate whether the one-loop corrections imparted by various scalars residing in the $45_{\hh}$ can modify the tree-level relations mentioned in Eq.~(\ref{eq:tree-yuk}). The couplings of various scalars stemming from $45_{\hh}$ with SM fermions at renormalisable level are shown below:
\beqa{\label{eq:45scalarscoupling}}
-\mathsf{L}^{45_{\hh}}_{\mathrm{Y}} & \supset & Y_{1\,AB}\Big( - \frac{1}{\sqrt{2}}\,\varepsilon_{ae}\,\varepsilon_{\beta\gamma\alpha}\,q^{e\beta\,T}_A\,\cc\,q^{b\gamma}_B\,\mathbb{T}^{\alpha\,a}_{b}+
 \sqrt{2}\,u^{C\,T}_{\alpha\,A}\,\cc\,e^C_B\,T^{\alpha} \nl \ad \frac{1}{\sqrt{{2}}}\,\varepsilon_{\beta\gamma\alpha}\,e^{C\,T}_A\,\cc\,q^{a\gamma}_B\,\Omega^{\dagger\,\alpha\beta}_a  - \frac{1}{\sqrt{2}}\,\varepsilon^{\gamma\lambda\theta}\,u^{C\,T}_A\,\cc\,u^C_{\lambda\,B}\,{\cal T}_{\theta}^{\dagger} \nl
\ad \sqrt{2}\,\varepsilon_{ab}\,q^{a\alpha\,T}_A\,\cc\,u^C_{\beta\,B}\,O^{\beta\,b}_{\alpha}- \frac{1}{2\sqrt{2}}\,\varepsilon_{ab}\,\varepsilon_{\gamma\alpha\beta}\,q^{a\gamma\,T}_A\,\cc\,q^{b\theta}_B\,S_{\theta}^{\alpha\beta}\Big)\nl
\ad Y_{2\,AB}\Big( \frac{1}{\sqrt{2}}\varepsilon^{\alpha\beta\rho}\,u^{C\,T}_{\rho\,A}\,\cc\,d^C_{\gamma\,B}\,S^{\dagger\gamma}_{\alpha\beta} + \frac{1}{\sqrt{2}}\,\varepsilon^{\alpha\beta\rho}\,u^{C\,T}_{\rho\,A}\,\cc\,d^C_{\alpha\,B}\,T^{\dagger}_{\beta}\nl
\ad \frac{1}{\sqrt{2}}\,\varepsilon^{\alpha\beta\rho}\,\varepsilon_{ab}\,u^{C\,T}_{\rho\,A}\,\cc\,l^b_B\,\Omega^a_{\alpha\beta} - \sqrt{2} q^{a\alpha\,T}_A\,\cc\,d^C_{\beta\,B}\,O^{\dagger\beta}_{\alpha\,a}  \nl 
\ad \sqrt{2}\,\varepsilon_{ab}\,q^{p\alpha\,T}_A\,\cc\,l^b_B\,\mathbb{T}^{\dagger\,a}_{p\alpha}\,-\,  \sqrt{2}\, e^{C\,T}_A\,\cc\,d^C_{\beta\,B}\,{\cal T}^{\beta}\, - \frac{1}{\sqrt{2}}\,\varepsilon_{ab}\,q^{a\alpha\,T}_A\,\cc\,l^b_B\,T^{\dagger}_{\alpha}\Big)\nl \hcn\,.
\eeqa
It is to be noted that the coupling of $T$ with quarks, i.e. $q\,q\,T$, is not allowed due to the antisymmetric nature of $Y_1$. Further, the scalar fields $T,\,{\cal T},\, \mathbb{T}$ exhibits diquark as well as leptoquark couplings, while all the other scalars either have diquark or leptoquark couplings. But, only scalar field $T$ and $\mathbb{T}$ can induce tree-level proton decay and all the other scalars exhibiting leptoquark couplings can induce proton decays at loop level.\\

The scalars contribute to one-loop Yukawa relations by inducing vertex corrections and correcting the external fermion leg. In addition to the corrections from the heavier scalar fields, the Yukawa vertices also receive contributions from the heavier gauge bosons. The couplings of heavy gauge boson~$(X)$ with the SM fermions is shown below~\cite{Buras:1977yy,Langacker:1980js}:
\beqa \label{eq:L_X}
-{\cal L}_{\rm G}^{(X)} &\supset& \frac{g_5}{\sqrt{2}} \overline{X}_\mu \left( \overline{d^C}_i \overline{\sigma}^\mu l_i - \overline{q}_i \overline{\sigma}^\mu u^C_i - \overline{e^C}_i \overline{\sigma}^\mu q_i \right) \hc\,, 
\eeqa
where $X$ has SM charge of $\big(3,2,-\frac{5}{6}\big)$, $g_5$ is the $SU(5)$ gauge coupling and $\sigma$ is the Pauli matrix.\\

The vertex corrections to the various Yukawa vertices are produced by heavier scalars and gauge bosons by propagating inside the loop. These vertex corrections can be computed from Eqs.~(\ref{eq:45scalarscoupling} and \ref{eq:L_X}) and are shown below:
\beqa{\label{eq:deltaf}}
\big(\delta Y_{u}\big)_{AB} \eq \frac{8}{\sqrt{6}}\, g_5^2 \big(Y_1\big)_{AB} f[M_X^2,0] -\sqrt{\frac{3}{2}}\,\big(Y_1^T\,Y_2^{*}\,Y_2^T\big)_{AB}\,f[M_{\Omega}^2,0]\nl \ad \sqrt{\frac{3}{2}}\,\left(\,Y_2\,Y_2^{\dagger}\,Y_1^T\,\right)_{AB}\,f[M_T^2,0] + \sqrt{6}\,\left(\,Y_2\,Y_2^{\dagger}\,Y_1\,\right)_{AB}\,f[M_O^2,0]\,,\nl
\big(\delta Y_d\big)_{AB} \eq \frac{2}{\sqrt{6}}\, g_5^2 \big(Y_2\big)_{AB} f[M_X^2,0] + 2\sqrt{6}\,\left(Y_1\,Y_1^{\dagger}\,Y_2\,\right)_{AB}\,f[M_O^2,0]\,,\nl
\big(\delta Y_e\big)_{AB} \eq -6\,\sqrt{\frac{3}{2}}\, g_5^2 \big(Y_2^T\big)_{AB} f[M_X^2,0] + 2\sqrt{6}\,\left( Y_2^T\,Y_1^{\dagger}\,Y_1^T\,\right)_{AB}\,f[M_\Omega^2,0]\nl \mi \sqrt{6} \left(Y_2^T\,Y_1^*\,Y_1\,\right)_{AB}\,f[M_T^2,0]\,.
\eeqa
where $(\delta Y)_{f}$ corresponds to the finite part of the vertex correction to the Yukawa vertex of fermion $f$ with SM Higgs. $f[M_i^2,0]$ is the loop integration factor, with its form provided in Appendix~(\ref{app:LF}),  $M_i$ being the masses of scalars listed in Tab.~(\ref{tab:scalars45H}) or the mass of the gauge boson~(X). These vertex corrections are a function of the renormalisation scale $\mu$ and are completely calculable in terms of the tree-level Yukawa couplings $Y_{1,2}$ and also depend upon the loop function involving the masses of heavy degrees of freedom. All the SM fields have been assumed massless while computing the vertex corrections to the Yukawa vertices. It can be noticed that $\delta\,Y_{d}$ and $\delta\,Y_e$ receive contributions from different scalars and hence are different.\\   

The wave function renormalisation factor of the (scalar or fermion) field \( f \) is computed by taking the derivative of the self-energy correction of field \( f \) with respect to the outgoing momentum and then setting the momentum to zero. The contribution to the wave function renormalisation due to various scalars and gauge bosons to the external leg of SM fermions and Higgs are shown below:
\beqa{\label{eq:kfs}}
\big(K_q\big)_{AB} \eq 3\,g_5^2\,\delta_{AB}\,h[M_X^2,0] - \Big( 6 h[M_O^2,0] + 4 h[M_{\mathbb{T}}^2,0] + 0.5 h[M_T^2,0] \Big)\, \big(Y_2\,Y_2^{\dagger}\big)_{AB} \nl \mi \Big( 2 h[M^2_{\mathbb{T}},0]+ 6 h[M_O^2,0] + \frac{3}{4}\,h[M_S^2,0]\Big)\big(\,  Y_1\,Y_1^{\dagger}\big)_{AB}\nl
\mi 2 h[M^2_\Omega,0]\,\big(Y_1^T\,Y_1^{*}\big)_{AB}\;,\nl
\big(K_{u^C}\big)_{AB} \eq 4\,g_5^2\,\delta_{AB}\,h[M_X^2,0] - \Big( 3 h[M_S^2,0] + 1.5 h[M_T^2, 0] + 2 h[M_\Omega^2,0] \Big)\, \big(Y_2\,Y_2^{\dagger}\big)_{AB} \nl
\mi \Big( 2 h[M_T^2,0] +  \frac{1}{2}\, h[M^2_{\cal{T}},0] + 6 h[M^2_O,0]\Big)\, \big(Y_1\,Y_1^{\dagger}\big)
_{AB}\;,\nl 
\big(K_{d^C}\big)_{AB} \eq 2\,g_5^2\,\delta_{AB}\,h[M_X^2,0] \nl\mi \Big( 6 h [M^2_S,0] + h[M_T^2,0] + 12 h[M^2_O,0] +  2\, h[M_{{\cal T}}^2,0]\Big)\,\big( Y_2^T\,Y_2^*\big)_{AB}\;, \nl
\big(K_l\big)_{AB} \eq 3\,g_5^2\,\delta_{AB}\,h[M_X^2,0] -\Big( 6 h[M^2_\Omega,0] + 6\, h[M_{\mathbb{T}}^2,0] + 1.5 h[M_T^2,0] \Big)\, \big(Y_2^T\,Y_2^*\big)_{AB}\;, \nl
\big(K_{e^C}\big)_{AB} \eq 6\,g_5^2\,\delta_{AB}\,h[M_X^2,0] \nl\mi \Big(  6\, h[M^2_{\cal T},0] + h[M_{\Omega}^2,0]\Big)\, \big(Y_1\,Y_1^{\dagger}\big)_{AB} - 6 \,h[M_T^2,0]\,\big(Y_1^T\,Y_1^*\big)_{AB}\;, \nl 
K_{H} \eq \frac{g_5^2}{2}\Big[2\,\left(f[M_X^2,M_T^2] + g[M_X^2,M_T^2]\right) + 4\,\left(f[M_X^2,M_{\mathbb{T}}^2] + g[M_X^2,M_{\mathbb{T}}^2]\right) \nl
\ad 4\,\left(f[M_X^2,M_{\cal T}^2] + g[M_X^2,M_{\cal T}^2]\right)\Big].
\eeqa
Different $K_f$ characterises the finite part of the wave function renormalisation factor corresponding to the field $f$. It receives corrections from heavy scalars and gauge bosons. $g$ and $h$ are different loop integration factors whose functional form is given in appendix~(\ref{app:LF}).\\

The following one-loop Yukawa relations at the renormalisation scale $\mu$ are obtained by substituting Eqs.~(\ref{eq:deltaf} and \ref{eq:kfs}) into Eq.~(\ref{eq:matchcond}):
\beqa{\label{eq:oneloopsu5}}
Y_{u} \eq \frac{2\,Y_1}{\sqrt{6}}\big(1 - 0.5\times K_H \big) + \delta_{Y_{u}} - 0.5 \times \frac{2}{\sqrt{6}}\times \Big(K_q^T\,Y_1 + Y_1\,K_{u^C}\Big)\;,\nl
Y_{d} \eq \frac{Y_2}{\sqrt{6}}\,\big(1 - 0.5\times K_H \big) + \delta_{Y_{d}} - \frac{0.5}{\sqrt{6}}  \times \Big(K_q^T\,Y_2 + Y_2\,K_{d^C}\Big)\;,\nl
Y_{e} \eq -\sqrt{\frac{3}{2}}\,Y_2^T\big(1 - 0.5 \times K_H \big) + \delta_{Y_{e}} - 0.5 \times \left( -\sqrt{\frac{3}{2}}\right) \times \Big(K_l^T\,Y_2^T + Y_2^T\,K_{e^C}\Big)\,.
\eeqa
The later part of this study is devoted to discussing the ability of Eq.~(\ref{eq:oneloopsu5}) to yield a valid charged fermion mass spectrum. This scenario, involving only \(45_\hh\) in the Yukawa sector, is labelled as $Model\,\, I$ for future reference.  

\subsection{ Model I-A: $SU(5)$ with $45_{\hh}$ and $15_{\hh}$}
\label{ssec:modelII}
The earlier subsection considered an \(SU(5)\) model with only \(45_{\hh}\) contributing to the Yukawa sector and examined its impacts on the Yukawa relations of charged fermions at one loop. This case is now extended by including \(15_{\hh}\) in the Yukawa sector to assess the model's ability to yield realistic charged and neutral fermion Yukawas and mixing angles and is termed as $model\,\, I-A$. The scalar fields residing in $15_{\hh}$ are listed in Tab.~(\ref{tab:scalarsin15}). In addition to generating neutrino masses, the scalars stemming from $15_{\hh}$ can also contribute to the correction of the Yukawa vertices.

\begin{table}[t]
    \centering
    \begin{tabular}{cc}\hline\hline
        ~~~~~~Notation~~~~~~ & ~~~~~~SM Charge~~~~~~ \\ \hline
        $t$ & $\big(1,3,1\big)$\\
        $\Delta$ & $\big(3,2,\frac{1}{6}\big)$ \\
        $\Sigma$ & $\big(6,1,-\frac{2}{3}\big)$ \\ \hline\hline

    \end{tabular}
    \caption{Scalar multiplets residing in $15_{\hh}$ dimensional irrep.}
    \label{tab:scalarsin15}
\end{table}

The inclusion of $15_{\hh}$ in the Yukawa sector modifies the Lagrangian as shown below:
\beqa{\label{eq:45and15}}
\mathsf{L}_Y \eq \mathsf{L}^{45_{\hh}}_Y + \mathsf{L}^{15_\hh}_Y\,,
\eeqa
where $\mathsf{L}^{45_\hh}_Y$ is written in Eq.~(\ref{eq:SU5Tree}). The decomposition of $15_{\hh}$ in its scalar submultiplets with canonically normalised kinetic term is as follows:
\beqa\label{eq:15deco}
15^{ab} \eq t^{ab},\hspace{1cm}15^{a\alpha}\;=\;\frac{1}{\sqrt{2}}\,\Delta^{a\alpha},\hspace{0.5cm}\text{and}\hspace{0.5cm}15^{\alpha\beta}\,=\,\Sigma^{\alpha\beta}\eeqa

The interaction of $15_{\hh}$ with $SU(5)$ fermion plets can be analogously decomposed into various effective vertices using the decomposition of $15_{\hh}$ and the resulting expression is shown below;
\beqa{\label{eq:SU15Htree}}
-\mathsf{L}_{\rm{Y}}^{15_{\hh}} \eq \big(Y_3\big)_{AB}\,\ff_A^T\,\cc\,\ff_B\,15_{\hh} \hc\nl
\eq Y_{3\,AB}\,\Big( \varepsilon_{am}\,\varepsilon_{bn}\,l^{m\,T}_A\,\cc\,l^n_B\,t^{ab} \,+\, \sqrt{2}\,\varepsilon_{ab}\,l^{b\,T}_A\,\cc\,d^C_{\alpha\,B}\,\Delta^{a\alpha}\,+\,d^{C\,T}_{\alpha\,A}\,\cc\,d^{C}_{\beta\,B}\,\Sigma^{\alpha\beta}\Big)\nl \hcn,
\eeqa
where $Y_3$ is symmetric in family labels. The interaction of SM fermions with the scalars originating from \(15_{\hh}\) is evaluated by decomposing \(15_{\hh}\) into its constituent scalars with canonically normalised kinetic terms and can be inferred from~\cite{Patel:2022wya}. The scalars residing in $15_{\hh}$ solely couple to fermions residing in $\ff$ dimensional irrep; consequently, they can only contribute to their wave-function renormalisation factor. Additionally, as there are no common scalars in $45_{\hh}$ and $15_{\hh}$, none of the scalars of $15_\hh$ can contribute to the vertex correction. The finite part of the corrections to the external leg of $d^C$ and $l$ due to $\Delta$, $\Sigma$ and $t$ are as follows:
\beqa{\label{eq:wfrfrom15}}
\big(k_{d^C}\big)_{AB} \eq -4\,h[M_{\Sigma}^2,0]\,\big(Y_3\,Y_3^{\dagger}\big)_{AB} - 4\,h[M_{\Delta}^2,0]\,\big(Y_3^T\,Y_3^*\big)_{AB}\nl
\big(k_l\big)_{AB} \eq -3\,h[M_{t}^2,0]\,\big(Y_3\,Y_3^{\dagger}\big)_{AB} - 6\,h[M_{\Delta}^2,0]\,\big(Y_3\,Y_3^{\dagger}\big)_{AB}\,,
\eeqa\\
where $h[m_1^2,m_2^2]$ is loop integration factor also used earlier. As \(\Sigma\) and \(t\) exhibit only diquark and dilepton couplings, respectively; these fields can specifically contribute to the external leg corrections of \(d^C\) and \(l\). In contrast, \(\Delta\), which exhibits leptoquark couplings, can correct the external legs of both \(d^C\) and \(l\) fermions.\\

The scalar irrep $15_\hh$ is conventionally known to contribute to the neutrino masses. The allowed $SU(5)$ invariant terms of the Lagrangian necessary for the generation of neutrino masses are as follows:
\beqa{\label{eq:lagNM}}
V_{\phi} \eq M^2_{45_\hh}\,45^{\dagger}_{\hh}\,45_{\hh} + M^2_{15_{\hh}}\,15^{\dagger}_{\hh}\,15_\hh + \Big(\lambda\,45_{\hh}\,45_{\hh}\,15^{\dagger}_{\hh} + \rho\, 45_{\hh}\,45_{\hh}\,45_{\hh}\,15_\hh \hc \Big)\,, \nl
\eeqa
where, $\lambda$ is a dimension-full trilinear coupling and is assumed to be close to the GUT scale while $\rho$ is the ${\cal{O}}\big(1\big)$ quartic coupling and has been set to unity. The scalar field \( t \) (cf. Table~\ref{tab:scalarsin15}) contributes to tree-level neutrino masses. At the loop level, contributions to neutrino masses arise from the scalar pairs \( T \)-\( \Delta \), associated with trilinear coupling $\lambda$ and \( \mathbb{T} \)-\( \Omega \) with the strength $\rho$~\cite{Dorsner:2016wpm,Dorsner:2017wwn}. The combined contribution to the neutrino mass from various sources up to one loop is as follows:
\beqa{\label{eq:SU5NM45n15}}
M_\nu \eq -2\,\lambda\,\frac{\langle H \rangle^2}{M_t^2}\,Y_3 + \frac{3}{2}\,\lambda\,\langle H \rangle^2\,\Big[ \big(Y_2^T\,Y_d^*\,Y_3^T\big) + \big(Y_2^T\,Y_d^*\,Y_3^T\big)^T \Big]\,\times\,p[M_T^2,M_\Delta^2]\nl
\ad 6\,\rho\,\langle t\rangle\,\langle H \rangle^2\,\Big[\big(Y_2^T\,Y_u^*\,Y_2\big) + \big(Y_2^T\,Y_u^*\,Y_2\big)^T\Big]\,\times\,p[M_{\mathbb{T}}^2,M_\Omega^2]. 
\eeqa
$\langle t \rangle $, $\langle H \rangle $ are vacuum expectation values of scalar fields $t$ and $H$ respectively. The definition of loop integration factor $p[m_1^2,m_2^2]$ has been relegated to the appendix~(\ref{app:LF}). The first term is the tree-level contribution, while the other two terms are the one-loop-level contribution. The loop-level expressions involve the contributions from one-loop generated $Y_d$ and $Y_u$. It is to be noted that $Y_{u}$ at tree-level is antisymmetric while the effectively generated $Y_u$ at one loop is not antisymmetric.\\

\subsection{Model II: $SU(5)$ with $5_{\hh}$ and $15_{\hh}$}
\label{ssec:modelIII}
Towards the end of this section, an alternative version of $SU(5)$ with a $5_{\hh}$-dimensional irrep contributing to the Yukawa sector is considered. This version of $SU(5)$ is known to produce generational mass degeneracy in the down-quark and charged lepton sectors, i.e., $Y_d = Y_e^T$~\cite{Langacker:1980js}. This generational mass degeneracy was addressed by extending the minimal \(SU(5)\) model with \(SU(5)\) fermion singlet(s), as presented in~\cite{Patel:2023gwt}. In the present study, an alternative approach to ameliorate the generational mass degeneracy is considered by extending the minimal \(SU(5)\) framework with a \(15_{\hh}\) irrep. Analogous to the previous subsections, one-loop corrections to various Yukawa vertices are computed, induced by the scalars in \(5_{\hh}\) and \(15_{\hh}\) as well as the heavy gauge bosons~$(X)$, and is labelled as $model-II$.\\

The interaction of $5_{\hh}$ and $15_{\hh}$ with the matter multiplets of $SU(5)$ is parameterised as shown below:
\beqa \label{eq:5+15}
-{\mathsf L}^{5_\hh + 15_\hh}_Y\eq \frac{1}{4}\big(Y_1\big)_{AB}\,\ft^T_A\,\cc\,\ft_B\,5_{\hh}\,+\; \sqrt{2}\,\big(Y_2\big)_{AB}\,\ft^T_{A}\,\cc\,\ff_{B}\,5^{\dagger}_{\hh}\,\nl \ad\, \big(Y_3\big)_{AB}\,\ff^T_A\,\cc\,\ff_B\,15_{\hh} \hc\,,
\eeqa
where $Y_1$ and $Y_3$ are symmetric in family indices. The decomposition of the term written in the last line of the Eq.~(\ref{eq:5+15}) is already given in Eq.~(\ref{eq:SU15Htree}) while the decomposition of the remaining terms is shown below:
\beqa \label{eq:SU5Hdecomposition}
-{\mathsf{L}}^{5_\hh}_{Y} \eq -Y_{1\,AB}\,\Big(u^{C\,T}_{\gamma\,A}\,\cc\,e^C_{B}\,+\, \frac{1}{2}\,\varepsilon_{\alpha\beta\gamma}\,\varepsilon_{ab}\,q^{a\alpha\,T}_{A}\,\cc\,q^{b\beta}_B\Big)T^{\gamma}\nl
\mi Y_{2\,AB}\,\Big(-\varepsilon_{ab}\,q^{a\gamma\,T}_A\,\cc\,l^b_B - \varepsilon^{\alpha\beta\gamma}\,u^C_{\alpha\,A}\,\cc\,d^C_{\beta\,B}\Big)T^{*}_{\gamma}\nl
\mi Y_{1\,AB}\,\varepsilon_{ab}\,q^{a\,\alpha}_A\,\cc\,u^C_{\alpha\,B}\,H^b\,-\, Y_{2\,AB}\,\Big( q^{a\alpha\,T}_A\,\cc\,d^C_{\alpha\,B} + e^{C\,T}_A\,\cc\,l^a_B\Big)\,H^{*}_a \hc\nl
\eeqa
It is evident from the above expression that the scalar field $T$ exhibits diquark as well as leptoquark coupling and hence can induce tree level proton decay.\\

The one-loop corrections to $Y_{u,d,e}$ in the presence of singlet fermions have been computed in~\cite{Patel:2023gwt}. These results are reproduced below for convenience, ignoring the singlet fermions' contributions. The finite part of the vertex corrections arising from the scalar field $T$ and the gauge bosons $X$ is as follows:
\beqa \label{eq:dY}
(\delta Y_u)_{AB} &=& 4 g_5^2 (Y_1)_{AB} f[M_X^2,0] + \left(Y_1 Y_2^* Y_2^T + Y_2 Y_2^\dagger Y_1^T \right)_{AB} f[M_T^2,0], \nonumber\\ 
(\delta Y_d)_{AB} &=& 2 g_5^2 (Y_2)_{AB} f[M_X^2,0]+\left(Y_1 Y_1^* Y_2 \right)_{AB} f[M_T^2,0]\,,\nonumber \\
(\delta Y_e)_{AB} &=& 6 g_5^2 (Y_2^T)_{AB} f[M_X^2,0]+ 3 \left(Y_2^T Y_1^* Y_1 \right)_{AB} f[M_T^2,0],
\eeqa
where $f$ is the loop integration factor given in appendix~(\ref{app:LF}). The first term is the contribution from heavy gauge boson while the other term is the contribution of the scalar field $T$. The wave function renormalisation factor of various fermion fields due to correction from $T$ and $X$ is also shown below:
\beqa \label{eq:K}
(K_{q})_{AB} &=& 3g_5^2 \delta_{AB} h[M_X^2,0] - \frac{1}{2}\left( Y_1^{*} Y_1^T +2  Y_2^* Y_2^{T}\right)_{AB} h[M_T^2,0], \nonumber\\
(K_{u^C})_{AB} &=& 4g_5^2 \delta_{AB} h[M_X^2,0] - \left(Y_1^* Y_1^T +2 Y_2^* Y_2^T\right)_{AB} h[M_T^2,0], \nonumber\\
(K_{d^C})_{AB} &=& 2g_5^2 \delta_{AB} h[M_X^2,0] - 2 \left(Y_2^\dagger Y_2\right)_{AB} h[M_T^2,0] \nonumber \\
(K_l)_{AB} &=& 3g_5^2 \delta_{AB} h[M_X^2,0] - 3 \left(Y_2^\dagger Y_2\right)_{AB} h[M_T^2,0], \nonumber\\
(K_{e^C})_{AB} &=& 6g_5^2 \delta_{AB} h[M_X^2,0] - 3 \left(Y_1^\dagger Y_1\right)_{AB} h[M_T^2,0], \nonumber\\
K_{H} &=& \frac{g_5^2}{2}\,\left(f[M_X^2,M_T^2] + g[M_X^2,M_T^2]\right), \eeqa 
where, again, $h$ and $g$ are loop integration factors can be inferred from the appendix~(\ref{app:LF}). The first term is the contribution of the gauge boson, while the second term is a contribution from the scalar field $T$. As there is no common multiplet in $5_\hh$ and $15_{\hh}$, scalars residing in $15_\hh$ can not contribute to the vertex correction of various Yukawa vertices. However, the scalars of $15_\hh$ can contribute to external leg correction of fermion living in $\ff$, which are $d^C$ and $l$, and such corrections have already been computed in Eq.~(\ref{eq:wfrfrom15}) and contribute to the $K_{d^C}$ and $K_l$. The one-loop matching condition, analogous to Eq.~(\ref{eq:oneloopsu5}), in this scenario can be computed by substituting Eqs.~(\ref{eq:dY}), (\ref{eq:K}), and (\ref{eq:wfrfrom15}) into Eq.~(\ref{eq:matchcond}) at a given renormalisation scale \(\mu\), yielding the following:
\beqa\label{eq:SU15atoneloop}
Y_{u} \eq Y_1\big(1 - 0.5\times K_H \big) + \delta Y_{u} - 0.5 \times  \Big(K_q^T\,Y_1 + Y_1\,K_{u^C}\Big)\;,\nl
Y_{d} \eq Y_2\,\big(1 - 0.5 \times K_H \big) + \delta Y_{d} - 0.5  \times \Big(K_q^T\,Y_2 + Y_2\,K_{d^C}\Big)\;,\nl
Y_{e} \eq Y_2^T\big(1 - 0.5 \times K_H \big) + \delta Y_{e} - 0.5 \times \Big(K_l^T\,Y_2^T + Y_2^T\,K_{e^C}\Big)\,.
\eeqa

Turning attention towards the contribution of $15_\hh$ towards the neutrino mass. The $SU(5)$ invariant allowed terms of the Lagrangian relevant for neutrino mass are as follows:
\beqa \label{eq:NM5+15}
V_{\phi} \eq M^2_{15_\hh}\,15_{\hh}^{\dagger}\,15_{\hh} + \Big(\lambda\,5_{\hh}\,5_{\hh}\,15^{\dagger}_{\hh} \hc \Big)\,,\eeqa
where the $\lambda$ is dimension-full cubic coupling and is assumed to be at the conventional GUT scale. The tree and loop contribution of $15_\hh$ in the neutrino mass can be analogously computed as shown in the subsection~(\ref{ssec:modelII}) and the result is shown below: 
\beqa\label{eq:NM15+5}
M_{\nu} \eq \langle H \rangle ^2\,\lambda\, \left( -\frac{2}{M_t^2}\,Y_3 - 3\sqrt{2}\,\Big[\big(Y_3\,Y_d^{\dagger}\,Y_2\big) + \big(Y_3\,Y_d^{\dagger}\,Y_2\big)^T \Big] \times\,p\big(M_\Delta^2,M_T^2\big)\right)\,,
\eeqa
where, the loop function $p[m_1^2,m_2^2]$ is defined in the appendix~(\ref{app:LF}). The first term in the above expression represents the tree-level contribution, while the second term is the loop-level contribution. The loop-level contribution is due to the propagation of a pair of scalars $T-\Delta$ inside the loop and includes the effectively generated $Y_d$ at one loop.

\section{Viability of Models}\label{sec:analysis}
In subsections~(\ref{ssec:modelI}), (\ref{ssec:modelII}), and (\ref{ssec:modelIII}), one-loop corrections to various Yukawa vertices are computed, with the latter two subsections also includes computations of neutrino mass within the considered framework. This section is dedicated to examining the viability of model-I and model-II in yielding the largest desired Yukawa values. The computed one-loop Yukawa relations depend on the Yukawa matrices~$(Y_{1,2,3})$, masses of the scalars and the heavy gauge boson, matching scale~$(\mu)$, and the \(SU(5)\) gauge coupling~$(g_5)$. The adopted strategy for the analysis fixes the masses of all scalars except two, along with the matching scale at the conventional GUT scale \((M_{\rm{GUT}} \sim 10^{16}\, \text{GeV})\) and a fixed gauge coupling value~$(g_{5}\,\sim\,0.5)$. The largest Yukawa entry of $Y_{1,2,3}$ is then varied within a specified range to reproduce the desired largest Yukawa entry, i.e. $Y_{u,d,e}$. The two-loop SM RGE extrapolated values require the largest entry of \(\big(Y_u\big)_{23} \sim 0.4\), while the desired values of $Y_d$ is $6\times 10^{-3}$ and for $Y_e$ is $9\times 10^{-3}$ at $\mu\,=\,$ \(10^{16}\) GeV~\cite{Patel:2023gwt}. The specified range of the elements of the Yukawa matrices \( Y_{1,2,3} \) is chosen such that it does not violate perturbative limits, i.e., \(\big(Y_{1,2,3}\big)_{ij} \leq (-3.5, 3.5)\), which is derived from \(2 \rightarrow 2\) tree-level scatterings at high energy limit~\cite{Allwicher:2021rtd}.
\subsection{Third Generattion Numerical Analysis: Model I}
\label{ssec:modelIanalysis}
Following form of $Y_{1}$ and $Y_2$ are chosen to check the ability of Eq.~(\ref{eq:oneloopsu5}) to address the inconsistencies mentioned in points~(\ref{pt:aa} and \ref{pt:bb}) simplifying the analysis:
\begin{itemize}
    \item $Y_1$ is assumed to be an antisymmetric matrix with complex entries, while $Y_2$ is taken as a real-diagonal matrix, as depicted below:
    \beqa{\label{eq:y1&y2form}}
    Y_1 \eq \begin{pmatrix} 
    0 & a & b \\
    -a & 0 & c \\
    -b & -c & 0
    \end{pmatrix}\,,\hspace{1cm}
    Y_2 \;=\; \begin{pmatrix}
        0 & 0 & 0 \\
        0 & 0 & 0 \\
        0 & 0 & d
    \end{pmatrix}\,,\eeqa
    where, $a,\,b,\,c$ are complex numbers while $d$ is purely real. The particular form of $Y_2$, in the above expression, has been specifically chosen to perform the third generation analysis.\\
\end{itemize} 
Substituting $Y_{1,2}$ given in Eq.~(\ref{eq:y1&y2form}) into Eq.~(\ref{eq:oneloopsu5}) results in the following functional form of $\big(Y_{u}\big)_{33}$, where the contribution of $K_H$ has been neglected;
\beqa{\label{eq:yu33}}
\big(Y_{u}\big)_{33} \eq 2\,i\,a\,Im\big(cb^*\big)\,\nl 
& \times& \Big( 2\,f[M_{\Omega}^2,0] + 2\,f[M_T^2,0] + \frac{1}{2}f[M_{\cal T}^2,0] + \frac{3}{4}\,f[M_S^2,0] + 2f[M_{\mathbb{T}}^2,0]\Big)\,,\eeqa
where $Im$ represents the imaginary part. It is to be noted that if any of $a$, $b$ or $c$ is $0$, then $\big(Y_{u}\big)_{33}$ is also $0$. Further, in the limit $\{a,b\}\to0$, $\big(Y_{u}\big)_{23}$ is obtained as follows:
\beqa{\label{eq:yu23}}
\big(Y_{u}\big)_{23} \eq c\,\sqrt{\frac{2}{3}} + \frac{g_5^2}{\sqrt{6}}\,\Big(8 h[M_X^2,0] - 7 f[M_X^2,0]\Big)\nl
\ad \sqrt{\frac{2}{3}}\,cd^2\,\Big(f[M_{\Omega}^2,0] + \frac{3}{2}\,f[M_S^2,0] + \frac{3}{4}\,f[M_T^2,0] + \frac{3}{2}\,h[M_{\Omega}^2,0] \Big)\nl
\ad \sqrt{\frac{2}{3}}\,|c|^2\,c\,\Big( f[M_\Omega^2,0] + 6\,f[M_O^2,0] + f[M_T^2,0] +  \frac{1}{4}\,f[M_{\cal T}^2,0] + \frac{3}{8}\,f[M_S^2,0] \nl \ad  f[M^2_{\mathbb{T}},0]\Big)\,.
\eeqa

Following comments and observations regarding the expressions given in Eqs.~(\ref{eq:yu33} and \ref{eq:yu23}) are in order:
\begin{enumerate}
    \item $Y_u$ is not antisymmetric:\\
Since the product of a diagonal matrix \( Y_2 \) and an antisymmetric matrix \( Y_1 \) contains only off-diagonal elements, \(\delta Y_{u}\) does not contribute to \(\big(Y_{u}\big)_{33}\). The only contribution to \(\big(Y_{u}\big)_{33}\) arises from a specific term in the wave function renormalisation correction, \(K_{q}\), which involves \(Y_1\,Y_1^{\dagger}\). This term is a Hermitian and determinant-less, and but includes diagonal elements. Although each individual contribution to \(Y_{u}\) is determinant-less (since \(Y_1\) itself is determinant-less), the combined effect of these contributions results in a \(Y_{u}\) that becomes determinant-full due to the influence of \(K_{q}\). Therefore, it is the contribution from \(K_{q}\) that renders \(Y_u\) a matrix with $det\neq 0$.
    \item $\big(Y_u\big)_{33} <$  $\big(Y_u\big)_{23}$. \\
The element $\big(Y_{u}\big)_{33}$ is generated at the loop level, whereas $\big(Y_{u}\big)_{23}$ is generated at the tree level; thus, $Y_{23}\;>\;Y_{33}$ and hence would be the largest element of $Y_{u}$ provided $c>b$.\\

\end{enumerate}

Having computed the largest entry in the up-quark sector, now down and charged lepton sectors are considered. In the limit $\{a,\,b\}\,\to\,0$, the expression of $\frac{\big(Y_d\big)_{33}}{\big(Y_e\big)_{33}}\;\equiv\;\frac{y_b}{y_\tau}$ turns out to be as follows: 
\beqa{\label{eq:yd-ye}}
\frac{y_b}{y_{\tau}}  & \sim & -\frac{1}{3} + \frac{4\,g_5^2}{3}\,\left(4\,f[M_T^2,0] - 5 h[M_X^2,0] \right) \nl
 \mi cc^*\Big[f[M_\Omega^2,0] + 3\,f[M_T^2,0] + \frac{2}{3}\big( f[M_{\cal T}^2,0] + f[M_{\mathbb{T}}^2,0] \big) + f[M_{O}^2,0] + \frac{1}{8}\,f[M_S^2,0]\nl
 \ad 4\,\big(\,h[M_\Omega^2,0] + h[M_O^2,0] - 2\,h[M_T^2,0]\big)\Big]\nl
 \mi 2\,d^2\,\Big[ \frac{3}{2}\,f[M_O^2,0] + \frac{1}{2}\,f[M_S^2,0] + \frac{1}{2}f[M_T^2,0] +  \frac{1}{6}\,f[M_{\cal T}^2,0] + \frac{7}{12}\,f[M_{\mathbb{T}}^2,0] \Big].
 \eeqa
As evident from the above expression that at the tree level $\frac{y_b}{y_{\tau}}$ equals $-\frac{1}{3}$, which it should be. At the one loop level this ratio receives corrections from various heavy degrees of freedom.\\ 

\begin{figure}[t]
\centering
\includegraphics[scale=0.65]{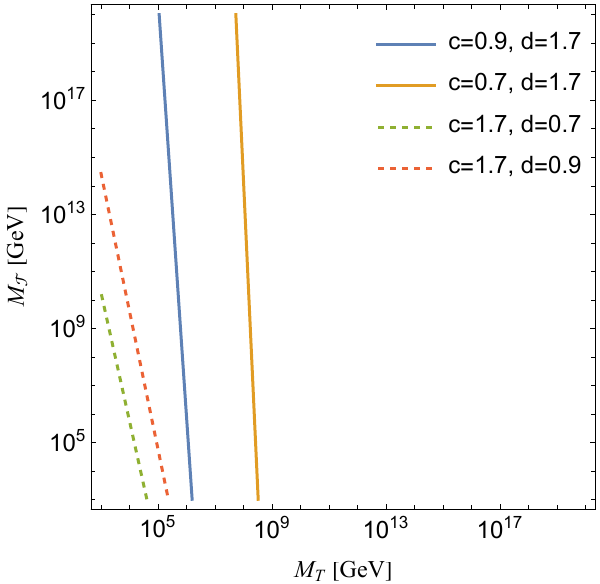}
\hspace{1cm} \includegraphics[scale=0.65]{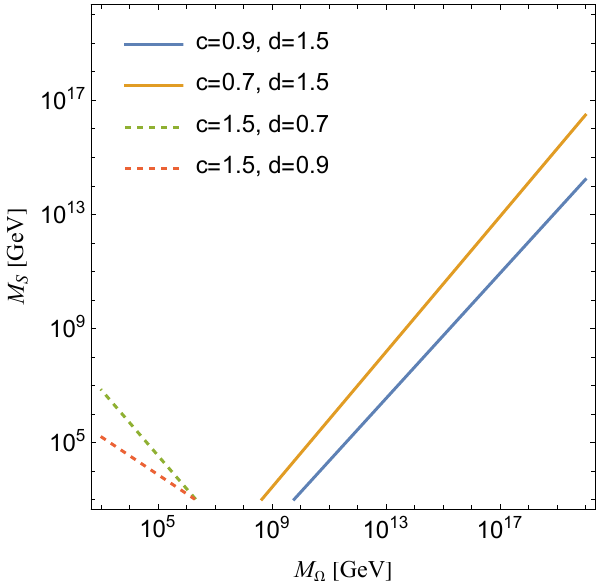}\\
\caption{Contours of $\big(Y_u\big)_{23} = 0.4$ for different values of $c$ and $d$ in $M_T\,-\,M_{\cal T}$~(left panel) and $M_\Omega - M_S$ planes~(right panel). All the other scalars and gauge bosons are assumed to be at the GUT scale.}
\label{fig:fig1}
\end{figure}


\begin{figure}[t]
\centering
\includegraphics[scale=0.65]{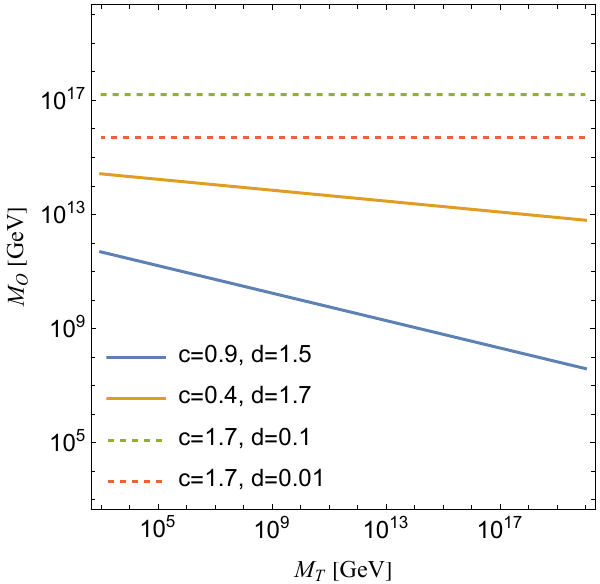}
\hspace{1cm} \includegraphics[scale=0.65]{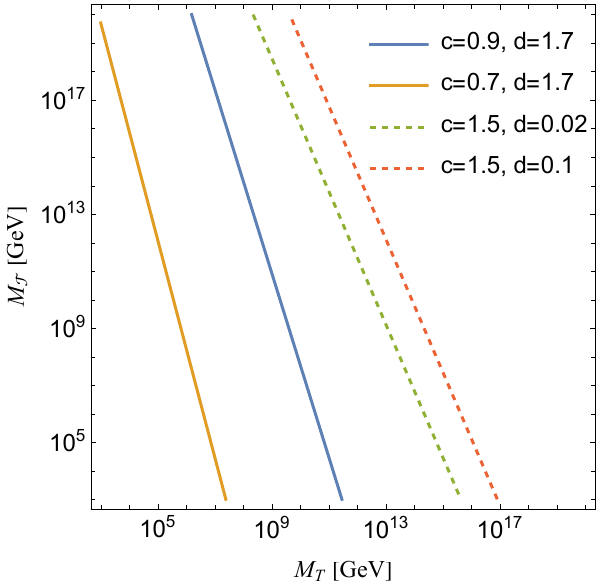}\\
\caption{Contours of $\big(Y_d\big)_{33} = 6\times 10^{-3}$ (left panel) and $\big(Y_e\big)_{33} = 9\times 10^{-3}$ (right panel) for different values of $c$ and $d$ in $M_T\,-\,M_O$ and $M_T\,-\,M_{\cal T}$ planes respectively. All the other scalars and gauge bosons are assumed to be at the GUT scale.}
\label{fig:fig2}
\end{figure}

Plots presented in Figs.~(\ref{fig:fig1}) examine the viability of Eq.~(\ref{eq:yu23}) in achieving specific values for \( \big(Y_u\big)_{23} \). The elements of \( Y_1 \) and \( Y_2 \), as defined in Eq.~(\ref{eq:y1&y2form}), are taken within the perturbative range \( (-3.5, 3.5) \), while the masses of the two relevant scalars are varied from 1 TeV to the Planck scale (\( M_p \sim 10^{19} \, \text{GeV} \)), keeping other heavy degrees of freedom at the conventional GUT scale \( M_{GUT} = 10^{16} \, \text{GeV} \). Similarly, the plots in Fig.~(\ref{fig:fig2}) examine the feasibility of Eqs.~(\ref{eq:yu23}) and~(\ref{eq:yd-ye}) in achieving specific values of \( y_b = 6 \times 10^{-3} \) and \( y_\tau = 9 \times 10^{-3} \) in the \( M_T-M_O \) and \( M_T-M_{\cal T} \) planes, respectively. All other parameters remain same with those in Fig.~(\ref{fig:fig1}). \\

These plots reveal that either \( c \) or \( d \) must exceed unity to achieve the desired values of \( \big(Y_u\big)_{23} \), $y_b$ and $y_\tau$. Furthermore, when \( c > d \), one of the scalar masses is more closer towards the matching scale (\( M_{GUT} \)). Conversely, for \( c < d \), the difference between the matching scale and the scalar mass increases. To summarise, in order to counteract suppression from loop factors, larger values of Yukawa couplings or a greater disparity between the matching scale and scalar mass are preferred. Thus far, the case with \( b = 0 \) and \( c \neq 0 \) has been considered in Eq.~(\ref{eq:y1&y2form}). If \( b \neq 0 \) and \( b > c \), then \( \big(Y_u\big)_{13} \) or \( \big(Y_u\big)_{31} \) would dominate, resulting in analogous outcomes.

\subsection{Third Generation Numerical Analysis: Model II}
\label{ssec:modelIIIanalysis}
The ability of the expressions in Eq.~(\ref{eq:SU15atoneloop}) together with (\ref{eq:wfrfrom15}) to yield specific values of the largest Yukawa couplings is examined in this subsection. For this purpose, the matrices \( Y_1 \), \( Y_2 \), and \( Y_3 \) are chosen in the following form:
\beqa\label{eq:y123form}
Y_1 \eq \begin{pmatrix}
    0 & 0 & 0\\
    0 & 0 & 0\\
    0 & 0 & a
    
\end{pmatrix}\,,\hspace{0.5cm}
Y_2 \;=\; \begin{pmatrix}
    0 & 0 & 0 \\
    0 & 0 & 0 \\
    0 & 0 & b
\end{pmatrix}\,,\;\rm{and}\;\;Y_3\;=\;\begin{pmatrix}
    0 & 0 & 0 \\
    0 & 0 & 0\\
    0 & 0 & c
\end{pmatrix}\,.\eeqa
where $b$ and $c$ can be purely real while $a$ is a complex number. Substituting the above form of $Y_{1,2,3}$ in Eqs.~(\ref{eq:SU15atoneloop}) and (\ref{eq:wfrfrom15}) leads to the following form of $\frac{y_b}{y_{\tau}}$:
\beqa\label{eq:yboverytaufor15}
\frac{y_b}{y_\tau} \;& \sim &\; 1 + 2g_5^2\,\big( f[M_X^2,0] - 2\,h[M_X^2,0] \big) - aa^*\,\big( \frac{5}{4}\,f[M_T^2,0] + 2\,h[M_T^2,0]\big)\nl
\ad \frac{b^2}{2}\,f[M_T^2,0] - c^2\,\big( f[M_\Delta^2,0] - f[M_S^2,0] + \frac{3}{2}\,f[M_t^2,0]\big)
\eeqa
The tree-level contribution to \(\frac{y_b}{y_\tau}\) is unity, deviating from unity due to contributions from various sectors. In this case, \(Y_1\) at tree level can be freely chosen to produce the desired value of \(\big(Y_u\big)_{33}\), hence its analysis is not done.\\
\begin{figure}[t]
\centering
\includegraphics[scale=0.65]{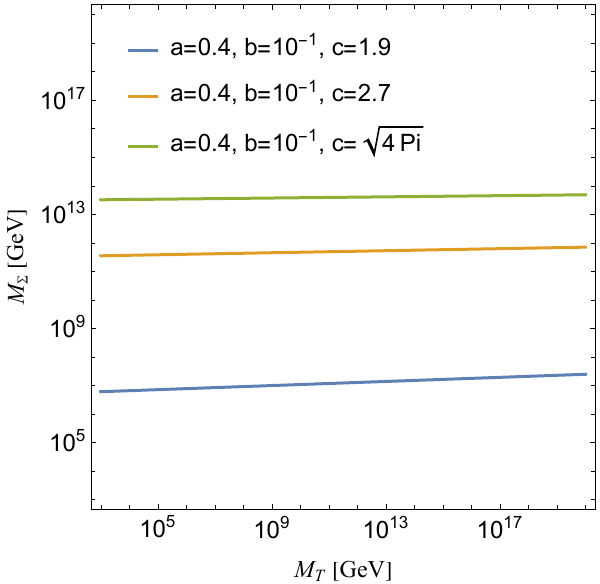}
\hspace{1cm} \includegraphics[scale=0.65]{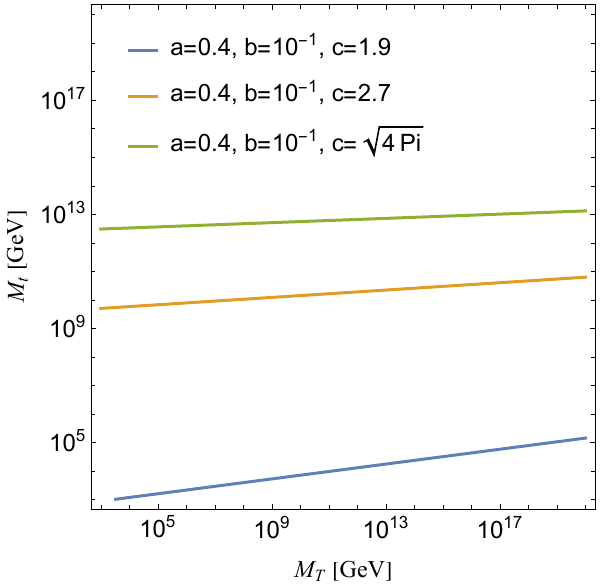}\\
\caption{Contours of $\big(Y_d\big)_{33} = 6\times 10^{-3}$ (left panel) and $\big(Y_e\big)_{33} = 9\times 10^{-3}$ (right panel) for different values of $c$ while holding $a$ and $b$ fixed, as they appear in Eq.~(\ref{eq:yboverytaufor15}),  in $M_T\,-\,M_\Sigma$ and $M_T\,-\,M_{t}$ planes respectively. All the other scalars and gauge bosons are assumed to be at the GUT scale.}
\label{fig:fig3}
\end{figure}

The plots in Fig.~(\ref{fig:fig3}) show contours of $y_b\,=\,6\times10^{-3}$ and $y_\tau\,=\,9\times 10^{-3}$ for different values of \(c\) in the \(M_T\)–\(M_\Sigma\) and $M_T - M_t$ planes, with \(a\) and \(b\) held constant at $0.4$ and $10^{-1}$, respectively. Additionally, the scalar field \(\Delta\), and the gauge boson \(X\) are also fixed at the conventional GUT scale, i.e. $M_X\,=\,M_\Delta\,=\,\mu\,=\,$ \(M_{\rm{GUT}} \sim 10^{16}\,\text{GeV}\). The contours in Fig.~(\ref{fig:fig3}) reveal that, as the magnitude of \(c\) increases, the difference between the matching scale and the mass of one of the scalars also increases. This arises due to the requirement for ample threshold corrections from the respective scalar. Therefore to achieve desired threshold corrections, either large couplings or a substantial difference between the mass of the scalar and the matching scale is preferred, which has also pointed out in the previous subsection. \\

\section{Quantitative Analysis}
\label{sec:solutions}
The discussions in section~(\ref{sec:analysis}) establish that the one-loop Yukawa relations computed for the different models in section~(\ref{sec:allmodels}) can, at least, reproduce the largest Yukawas of \( Y_u \), \( Y_d \), and \( Y_e \). This section presents the three-generation \(\chi^2\) analysis for the different models considered. The \(\chi^2\) definition and optimisation procedure can be inferred from~\cite{Joshipura:2011nn, Mummidi:2021anm, Patel:2023gwt} and is reproduced below for convenience;
\beqa\label{eq:chisqdef}
\chi^2 \eq  \sum_{i} \left(\frac{O_{i\,\rm{theo}} - O_{i\,\rm{exp}}}{\sigma_{i\,\rm{exp}}}\right)^2\,,
\eeqa
where, the theoretical prediction for each observable is denoted by \( O_{i,\mathrm{theo}} \), while \( O_{i,\mathrm{exp}} \) represents the corresponding experimentally measured value, where \( i \) runs over all observables in the dataset. The experimental uncertainty associated with each measurement is given by \( \sigma_i \). As, all the $O_{i\,\rm{theo}}$ are computed at the GUT scale, $O_{i\,\rm{exp}}$ will also be taken at GUT scale. The GUT scale values of the different $O_{i}$'s are done performing two SM renormalisation group evolution from $M_Z$ to the GUT scale~$(M_{\rm{GUT}} \sim 10^{16}\,\rm{GeV})$~(cf. footnote\footnote{ The calculation starts with a basis at $M_Z$ where $Y_{u,e} = \sfrac{\rm{Diag}\big(M_{u,e}\big)}{v}$, $Y_d = \sfrac{V_{\rm{CKM}} M_d}{v}$. Here $M_{u,d,e}$ are the diagonal mass matrices of charged fermions at $M_Z$ and $v$ is the vacuum expectation value of SM Higgs. These values are used as input for the 2-loop SM RGE equations. The running performed using the~\cite{Sartore:2020gou} gives the observables at the GUT scale. After diagonalisation, the GUT-scale observables are obtained and results are listed in Tab.~(\ref{tab:sol1}).})). While the experimental uncertainties of various observables differ and are only known at the low-energy scale, a conservative estimate has been adopted for these observables at the GUT scale. For lighter quarks, an experimental uncertainty of $\pm\SI{30}{\percent}$ is assumed, while for other observables the uncertainty is set to $\pm\SI{10}{\percent}$. Although leptonic parameters are known with high precision at low energies~\cite{ParticleDataGroup:2024cfk}, a conservative uncertainty of $\pm\SI{10}{\percent}$ is imposed at the GUT scale due to several factors: (1) unaccounted scalar  contributions to RGE evolution, (2) uncertainties in scalar masses, and (3) uncertainty in the matching scale (4) next-to-next leading order threshold corrections. The ambiguity in the known physics at the GUT scale is the primary reason of the increased uncertainty~(cf.~\cite{Babu:2016bmy,Dueck:2013gca} for fits having different uncertainty level for the leptonic spectrum).

\subsection{Model I}
\label{ssec:solmodelI}
The viability of \(Y_u\), \(Y_d\), and \(Y_e\) given in Eq.~(\ref{eq:oneloopsu5}) to yield realistic low-energy observables: Yukawa couplings of charged fermions and the quark mixing matrix—is assessed through \(\chi^2\) optimisation. The \(\chi^2\) definition includes six quark Yukawas, three charged lepton Yukawas, and four quark mixing parameters. The two-loop RGE evolved values of the SM Yukawas of the charged fermions at the GUT scale~$(M_{\rm{GUT}}\sim10^{16}\,\rm{GeV})$ has been sourced from~\cite{Patel:2023gwt} and is listed as $O_{\rm exp}$ in Tab.~(\ref{tab:sol1}). \\

The basis given in Eq.~(\ref{eq:y1&y2form}) for \(Y_1\) and \(Y_2\) is chosen again, making \(Y_1\) a complex antisymmetric matrix and \(Y_2\) a real diagonal matrix, providing a total of fifteen parameters, which are nine Yukawa couplings and six scalar masses, to be fitted in order to reproduce thirteen aforementioned observables. Additionally, the \(SU(5)\) gauge coupling is taken to be \(g_{5}\,=\,0.524\), and the matching scale~$(\mu)$ is set at the conventional GUT scale~$(M_{\rm{GUT}})$, which is also chosen as the mass of the heavy gauge bosons~$(M_X)$: $\mu\,=\,M_{\rm{GUT}}\,=\,10^{16}\,\rm{GeV}\,=\,M_X$. The mass of the heavy gauge boson~$(M_X)$ has been chosen to comply with the proton decay bounds. The Yukawa couplings have been restricted to lie within the perturbative regime i.e. $\big|\big(Y_{1,2}\big)_{ij}\big| \leq 4\pi$. The values of various parameters obtained in the minimum $\chi^2$ analysis are provided in Tab.~(\ref{tab:sol1}) and is labelled as solution~(I).

\begin{table*}[t!]
	\begin{center} 
		\begin{math} 
			\begin{tabular}{cccccc}
				\hline
				\hline 
      \multirow{2}{*}{~~Observable~~} & \multirow{2}{*}{ \hspace*{0.8cm}  $O_{\rm exp}$ \hspace*{0.8cm} } & 
      \multicolumn{2}{c}{\bf \hspace*{0.8cm}  Solution I \hspace*{0.8cm} } &
      \multicolumn{2}{c}{\bf \hspace*{0.8cm}  Solution IA \hspace*{0.8cm} }\\
				 &   & $O_{\rm th}$ & pull & $O_{\rm th}$ & pull \\
				\hline
				$y_u$  & $2.81 \times 10^{-6}$ & $2.77 \times10^{-6}$& $ 0$ & $2.80\times 10^{-6}$ & $0$ \\
				$y_c$   & $1.42\times10^{-3}$ & $1.42\times10^{-3}$& $0$ & $1.42 \times 10^{-3}$ & $0$\\
				$y_t$   & $4.27 \times 10^{-1} $ & $4.28\times10^{-1}$& $0.1$ & $4.25 \times 10^{-1}$& $ 0$\\
				$y_d$  & $ 6.14 \times 10^{-6} $ & $ 3.77\times 10^{-6}$ & $-1.2$ & $5.45 \times 10^{-6}$ & $-0.4$\\
				$y_s$   & $1.25 \times10^{-4}$ & $1.78\times 10^{-4}$ & $1.4$ & $1.28\times 10 ^{-4}$ & $0.1$\\
				$y_b$   & $5.80 \times10^{-3}$ & $5.80\times 10^{-3}$& $0$ & $5.39\times 10^{-3}  $ & $-0.7$\\
				$y_e$   & $2.75 \times 10^{-6}$ & $2.81 \times 10^{-6}$& $0.2$ & $2.74 \times 10^{-6}$ & $0$\\
				$y_{\mu}$  &$5.72
                \times10^{-4}$ & $5.30 \times 10^{-4}$ & $-0.7$ & $5.93 \times 10^{-4}$& $ 0.4 $\\
				$y_{\tau}$   & $9.68 \times10^{-3}$ & $9.68 \times 10^{-3}$ & $0$ & $1.01\times 10^{-2}$ & $ 0.4$\\
				$|V_{us}|$ & $0.2286$ & $0.2160$& $-0.6$ & $0.2286$ & $ 0$\\
				$|V_{cb}|$ & $0.0457$ & $0.0438$& $-0.4$ & $0.0443$ & $ -0.3$\\
				$|V_{ub}|$ & $0.0042$ & $0.0044$ & $0.4$ & $0.0043$ & $ 0$ \\
				$\sin\delta_{\rm CKM}$ & $0.78$ & $0.88$ & $ 1.3$ & $0.78$ & $ 0$  \\
				$\Delta m^2_{\text{sol}}~ [{\rm eV}^2]$ & $7.49\times10^{-5}$& -- & -- & $7.74\times 10^{-5}$ & $ -0.1 $\\
				$\Delta m^2_{\text{atm}}~ [{\rm eV}^2]$ & $2.534\times10^{-3}$& -- & -- & $2.686 \times 10^{-3}$ & $ 0.1$ \\
				$\sin^2 \theta _{12}$ & $0.307$ & -- & -- & $0.327$ & $ 0.3$\\
				$\sin^2 \theta _{23}$ & $0.561$ & -- & -- & $0.526$ & $ 0.5$  \\
				$\sin^2 \theta _{13}$ & $0.02195$ & --& -- & $0.02249$ & $ 0$ \\
    \hline
                $\chi^2_{\rm min}$    & & & $6.6$ & & $2.0$  \\
    \hline
                $M_T$ [GeV]& & $1.002 \times 10^{14}$& & $5.973\times10^{15} $ &\\
                $M_{\cal T}$ [GeV]& & $ 1.060 \times 10^{14}$ & & $1.201 \times 10^{15}$&\\
                $M_{\Omega}$ [GeV] & & $8.115 \times 10^{9}$& & $6.411 \times 10^{15}$ &\\
                $M_{\mathbb{T}}$ [GeV] & & $1.001\times 10^{14}$& & $1.133\times 10^{15}$ & \\
             $M_{S}$ [GeV] & & $3.122\times 10^{8}$ &  & $2.034\times10^5$  \\
             $M_{O}$ [GeV] & & $3.204\times 10^{7}$ & &$5.064\times10^{15}$ &  \\
             $M_{\Delta}$ [GeV] & &$-$ & & $4.899\times10^{13}$&  \\
             $M_{\Sigma}$ [GeV] & &$-$ & & $1.928\times10^{11}$  \\
             $M_{t}$ [GeV] & & $-$ & &$2.565\times10^{15}$ & \\
             $\lambda$ [GeV] & & $-$ & & $1.089\times10^{16}$ & \\
             $\langle t \rangle $ [GeV] &  & $-$ & & $1.000$ & \\
    \hline
				\hline 
			\end{tabular}
		\end{math}
	\end{center}
	\caption{The benchmark best-fit values obtained for solutions (I) and (II). In addition to the charged fermion sector observable, $\chi^2$ also includes observables of neutral leptons in the case of solution (II).  The extrapolated values at the scale of $\mu=10^{16}$ GeV are provided along with the values reproduced through $\chi^2$ minimisation and the corresponding pulls. The fitted value of mass of the various scalars, cubic coupling $\lambda$ and vev of the scalar field $t$ are also specified at the bottom of the table. } 
	\label{tab:sol1} 
\end{table*}

The fitted values of $Y_{1,2}$ corresponding to entries of solution~I in Tab.~(\ref{tab:sol1}) are as follows:
\beqa\label{eq:fittedy1y2}
Y_1 \eq \begin{small} \begin{pmatrix}
    0 & \big(0.895 + i\,1.337\big)\times 10^{-4} & -0.585455 - i\,0.834302\\
    \big(-0.895 - i\,1.337\big)\times 10^{-4} & 0 & -0.0298835 - i\,0.0452801 \\
    0.585455 + i\,0.834302 & 0.0298835 + i\,0.0452801 & 0
    
\end{pmatrix}\,\end{small},\nl
Y_2 \eq \begin{pmatrix}
    -2.6 \times 10^{-6} & 0 & 0 \\
    0 & 0.0004304 & 0 \\
    0 & 0 & -1.51563
\end{pmatrix}\,.
\eeqa

It is evident from the Yukawa relations derived in Eq.~(\ref{eq:oneloopsu5}) that $Y_{1,2}$ collectively contribute to $Y_{u,d,e}$ at one loop. Moreover, as seen from the fitted values of $Y_{1,2}$ given in Eq.~(\ref{eq:fittedy1y2}), the largest entry of $Y_2$ is comparable to the largest entry of $Y_1$. However, the largest entry of $Y_u$ must be more than two orders of magnitude greater than the largest entry of $Y_{d,e}$ in order to reproduce the correct order of the Yukawa couplings. Therefore, a cancellation occurs between the tree-level Yukawa coupling of $Y_{d,e}$ and the loop-corrected Yukawa, resulting in smaller Yukawa values, as show below;
\beqa\label{eq:delyd form}
\left|\frac{1}{\sqrt{6}} Y_{2}\right| \eq \begin{pmatrix}
    -1.06 \times 10^{-6} & 0 & 0 \\
    0 & 1.75\times 10^{-4} & 0 \\
    0 & 0 & 0.615\end{pmatrix} \\
    \big|\delta Y_{d}\big| &=& \begin{pmatrix}
    3.37 \times 10^{-6} & 2.96 \times 10^{-5} & 1.65\times10^{-5} \\
    1.79\times 10^{-7} & 2.19 \times 10^{-6} & 3.09\times10^{-4}\\
    2.82\times 10^{-11} & 8.78 \times 10^{-8} & 1.98
    \end{pmatrix}\,,
\eeqa\

\begin{eqnarray}\label{eq:kq form}
 \left| \frac{0.5}{\sqrt{6}}  \times \Big(K_q^T\,Y_2 + Y_2\,K_{d^C}\Big) \right| &=& \begin{pmatrix}
    5.93 \times 10^{-7} & 5.24 \times 10^{-6} & 2.91\times10^{-6} \\
    3.16\times 10^{-8} & 8.79 \times 10^{-8} & 5.47\times10^{-5}\\
    4.99\times 10^{-12} & 1.55 \times 10^{-8} & 2.59
    \end{pmatrix}\,.\nonumber\\
    \Rightarrow  \left|\frac{1}{\sqrt{6}} Y_{2}\right|_{33} + \left| \delta Y_d \right|_{33} &-&  \left| \frac{0.5}{\sqrt{6}}  \times \Big(K_q^T\,Y_2 + Y_2\,K_{d^C}\Big) \right|_{33} \approx 6 \times 10^{-3}
\end{eqnarray}
As shown above, the Yukawa correction involves the product of three Yukawa matrices. Since the largest entry of each matrix, \( Y_{1,2} \), is \({ \cal O}(1)\), and the loop function can contribute at most one order less than the largest entry, the largest entry of the loop-corrected Yukawa is also \({ \cal O}(1)\), due to which the cancellation happens leading to small Yukawa.  \\

The \(\chi^2\) analysis indicates that all the thirteen known parameters can be nicely fitted within the allowed range, resulting in a \(\chi^2\) value of $6.6$. As evident from the values of $Y_{1,2}$ given in Eq.~(\ref{eq:fittedy1y2}), the Yukawa entries remain well within perturbativity constraints. The masses of the three scalars \(M_{\Omega}\), \(M_{S}\), and \(M_{O}\) differ maximum from the matching scale, contributing ample threshold corrections while other scalars are closer towards the matching scale. However, this substantial mass splitting among the scalars challenges the Extended Survival Hypothesis.  As evident from this case, the splitting among the masses of different scalars belonging to the same irrep~(i.e. $45_\hh$) significantly contributes to altering the tree-level Yukawa relations. But, this requires an alternate mechanism that creates substantial hierarchies among different scalar fields within the same multiplet.\\ 

Before discussing the three-generation \(\chi^2\) analysis of models-I-A and II, constraints on the scalar fields $T$ and \(\mathbb{T}\) from the proton decays are presented in the following subsection.
 
\subsubsection{Constraints on scalar fields $T$ and $\mathbb{T}$ from Proton Decay}
\label{sssec:protondecay}
As evident from Eq.~(\ref{eq:45scalarscoupling}), the scalar fields \( T \), \( {\cal T} \), and \( \mathbb{T} \) exhibit diquark and leptoquark couplings and can thus induce proton decay. The scalar field \( {\cal T} \), however, couples to up quarks with flavour-antisymmetric couplings, resulting in proton decay mediated by \( {\cal T} \) occurring only at the loop level. In contrast, \( T \) and \( \mathbb{T} \) can induce tree-level proton decays. Additionally, the scalar field \( \Omega \) exhibits leptoquark couplings that can mix with the diquark coupling of \( \mathbb{T} \), thereby inducing proton decay. Such mixing occurs when the SM Higgs acquires a $vev$, leading to the mixing of states with identical electric charges, particularly the \(\sfrac{2}{3}\)-charged state in this case. However, this mixing is ruled out if there are only \( 24_{\hh} \) and \( 45_{\hh} \) irreps present in the model as none of the combinations $45_{\hh}\,45_\hh\,45_\hh$, $45_\hh\,45_\hh\,45^{\dagger}_\hh$, $45_\hh\,45^{\dagger}_{\hh}45^{\dagger}_\hh$, $24_\hh\,45_{\hh}\,45_\hh\,45_\hh$, $24_\hh\,45_\hh\,45_\hh\,45^{\dagger}_\hh$ and $24_\hh\,45_\hh\,45^{\dagger}_{\hh}45^{\dagger}_\hh$ are $SU(5)$ invariant . Such mixing can arise only if the scalar sector also includes the \( 15_{\hh} \)-dimensional irrep which is $15_\hh\,45_\hh\,45_\hh\,45_\hh$, as in the case of \( \text{model-IA} \). \\

As a result, only tree-level proton decays mediated by \( T \) and \( \mathbb{T} \) have been considered. Integrating out the degrees of freedom of $T$ and $\mathbb{T}$ results in the following dimension-six four-fermion operators:
\beqa\label{eq:PD}
\mathsf{L}^{\slashed{B}}_{\rm{eff}} \eq h_1[A,B,C,D]\,\varepsilon_{\alpha\beta\gamma}\,\left(u^{\alpha\,T}_{A}\,\cc\,d^{\beta}_{B}\right)\,\left(u^{\gamma\,T}_C\,\cc\,e_D\right)\nl
\ad h_2[A,B,C,D]\,\varepsilon_{\alpha\beta\gamma}\,\left(u^{\alpha\,T}_{A}\,\cc\,d^{\beta}_{B}\right)\,\left(d^{\gamma\,T}_C\,\cc\,\nu_D\right)  \nl
\ad h_{3}[A,B,C,D]\,\varepsilon_{\alpha\beta\gamma}\, \left(u^C_{\alpha\,A}\,\cc\,d^C_{\beta\,B}\right)^*\,\left(u^{\gamma\,T}_A\,\cc\,e_B\right)\nl
\ad h_{4}[A,B,C,D]\,\varepsilon_{\alpha\beta\gamma}\, \left(u^C_{\alpha\,A}\,\cc\,d^C_{\beta\,B}\right)^*\,\left(d^{\gamma\,T}_A\,\cc\,\nu_B\right)\hc \nl 
\eq h_1[A,B,C,D]\,\varepsilon_{\alpha\beta\gamma}\,\left(\overline{u^{\alpha\,C}_L}_{A}\,d^{\beta}_{L\,B}\right)\,\left(\overline{u^{\gamma\,C}_L}_C\,e_{L\,D}\right)\nl
\ad h_2[A,B,C,D]\,\varepsilon_{\alpha\beta\gamma}\,\left(\overline{u^{\alpha\,C}_L}_{A}\,d^{\beta}_{L\,B}\right)\,\left(\overline{d^{\gamma\,C}_L}_C\,\nu_{L\,D}\right)\,\nl
\ad h_3[A,B,C,D]\,\varepsilon_{\alpha\beta\gamma}\,\left(\overline{u^{\alpha\,C}_R}_{A}\,d^{\beta}_{R\,B}\right)\,\left(\overline{u^{\gamma\,C}_L}_C\,e_{L\,D}\right)\nl
\ad h_4[A,B,C,D]\,\varepsilon_{\alpha\beta\gamma}\,\left(\overline{u^{\alpha\,C}_R}_{A}\,d^{\beta}_{R\,B}\right)\,\left(\overline{d^{\gamma\,C}_L}_C\,\nu_{L\,D}\right)\,\hc,
\eeqa
where $h_{1,2,3,4}$ are effective strengths of the operators and depend upon particular flavour indices. The last line of the above expression represents the baryon number violating operator in the conventional form where $L$ and $R$ denotes the chirality of the respective fermion. These operators violates the baryon number and induces proton decay modes having $B-L\,=\,0$. The functional form of $h_{1,2}$ in the mass basis in terms of coupling $Y_{1,2}$ are shown below:
\beqa\label{eq:effectivestrenth}
h_{1}[A,B,C,D] \eq \frac{1}{M_{\mathbb{T}}^2}\,\left[\Big(U^T\,Y_1\,D\Big)_{BA}\,\Big(D^T\,Y_2\,E\Big)_{CD} + 2\,\Big(U^T\,Y_1\,U\Big)_{CB}\,\Big(D^T\,Y_2\,E\Big)_{AD}  \right]\nl
h_{2}[A,B,C,D] \eq \frac{1}{M_{\mathbb{T}}^2}\,\left[\Big(U^T\,Y_1\,D\Big)_{BA}\,\Big(D^T\,Y_2\,N\Big)_{CD} + 2\,\Big(D^T\,Y_1\,D\Big)_{AC}\,\Big(U^T\,Y_2\,N\Big)_{BD}  \right]\nl
h_{3}[A,B,C,D] \eq \frac{1}{2\,M_{T}^2}\,\left[\Big(U^{C\,\dagger}\,Y_2^{\dagger}\,D^{C\,*}\Big)_{AB}\,\Big(U^T\,Y_2\,E\Big)_{CD} \right]\nl
h_{4}[A,B,C,D] \eq \frac{-1}{2\,M_{T}^2}\,\left[\Big(U^{C\,\dagger}\,Y_2^{\dagger}\,D^{C\,*}\Big)_{AB}\,\Big(D^T\,Y_2\,N\Big)_{CD} \right]\,,
\eeqa
where $U,U^C,D,D^C,E,N$ are the unitary matrices which diagonalises the $Y_{u}$, $Y_d$, $Y_e$ and $M_{\nu}$ respectively and their evaluation can be conferred from~\cite{Patel:2022wya}. The effective strengths are suppressed by the quadratic mass of the scalar fields $T$ and $\mathbb{T}$. Two-component Fierz transformation rules for Weyl fermions have been used in order to evaluate the effective strengths of the operators~\cite{Dreiner:2008tw}.

\begin{table}[t]
\begin{center}
\begin{tabular}{lcc} 
\hline\hline
~~~~~~~~~~Branching ratio [\%]~~~& ~~~~~~~~$T$~~~~~~~~&~~~~~~~~$\mathbb{T}$~~~~~~~~\\

\hline
~~~~~~~~\hspace{1cm}${\rm BR}[p\to \bar{\nu}\,\, \pi^+]$~~~~~~~  & $99.9$ & $<1$ \\
~~~~~~~~\hspace{1cm}${\rm BR}[p\to \bar{\nu} K^+]$~~~~~~~  & $<1$ &$99.8$ \\
${\rm BR}[p\to e^+ \pi^0,\, \mu^{+} \pi^0, e^+ K^0, \mu^+ K^0 ]$ & $< 1$ & $<1$ \\
\hline\hline
\end{tabular}
\end{center}
\caption{Proton decay branching fraction estimated for the solution~(I) of Tab.~(\ref{tab:sol1}) induced by the scalar fields $T$ and $\mathbb{T}$. Only the contribution of a single scalar at a time is considered when computing the bounds on the mediators.}
\label{tab:PD}
\end{table}

The scalar field $\mathbb{T}$ couples only to left-chiral fermions, consequently the proton decay operator it induces only involves left-chiral fermions, while the scalar field $T$ couples to right and left chiral fermions. To evaluate their contribution to proton decay, the contribution from either $T$ or $\mathbb{T}$ is considered at a time, assuming the other scalar is significantly heavier than the one under consideration. The expressions of leading two body proton decay modes can be inferred from~\cite{Nath:2006ut} while the values of various parameters entering in the decay width expression and the procedure of evaluating the decay widths can be referenced from~\cite{Patel:2022wya}. Substituting the fitted values of $Y_1$ and $Y_2$ from Eq.~(\ref{eq:fittedy1y2}) into Eq.~(\ref{eq:effectivestrenth}) and using matrices for $U, D$, $U^C$, $D^C$ and $E$, one can compute the branching ratios of leading two body proton decay modes and are given in Tab.~(\ref{tab:PD}). Additionally, as neutrinos are massless in the setting of solution~I, consequently $N$ matrix appearing in Eq.~(\ref{eq:effectivestrenth}) is taken to be an identity matrix.\\

When the contribution to proton decay arises solely from $T$, the leading decay mode is $p \to \overline{\nu}\,\pi^+$, and is due to the diagonal nature of $Y_2$. The lower bound on the mass of the scalar field $T$, ensuring consistency with the experimental lower limit on proton decay mode~$p \to \overline{\nu}\,\pi^+$~\cite{Super-Kamiokande:2013rwg}, is given as follows:
\beqa\label{eq:LBonMT}
\tau/{\rm BR}[p \to \overline{\nu}\, \pi^+] &=& 3.9 \times 10^{32}\,{\rm yrs}\,\times \left(\frac{M_{T}}{1.7 \times 10^{13}\,{\rm GeV}} \right)^4\,.\eeqa

It is also evident from the Tab.~(\ref{tab:PD}) that the leading branching mode is $p\to \overline{\nu}\,K^+$ when proton decay is only induced with $\mathbb{T}$ and is due to summing over all the flavours of neutrinos and presence of contribution of $Y_1$ matrix. The lower bound on the mass of $\mathbb{T}$ from  the current lower bound on the mode $p\to \overline{\nu}\,K^+$~\cite{Super-Kamiokande:2014otb} is as follows:
\beqa\label{eq:LBonMT33}
\tau/{\rm BR}[p \to \overline{\nu}\, K^+] &=& 5.9 \times 10^{33}\,{\rm yrs}\,\times \left(\frac{M_{\mathbb{T}}}{1.0 \times 10^{14}\,{\rm GeV}} \right)^4\,.\eeqa
The fitted value of $M_{T}$ and $M_{\mathbb{T}}$ in solution~(I) of Tab.~(\ref{tab:sol1}) is greater than the lower bound on it from the proton decays. Moreover, the fitted value of \( {\cal T} \), being only two orders of magnitude less than the GUT scale, also adheres to the constraints from loop-induced proton decays.

\subsubsection{Effect of lightest scalar in Model-I}

\begin{figure}[t!]
    \centering
    \includegraphics[width=1\linewidth]{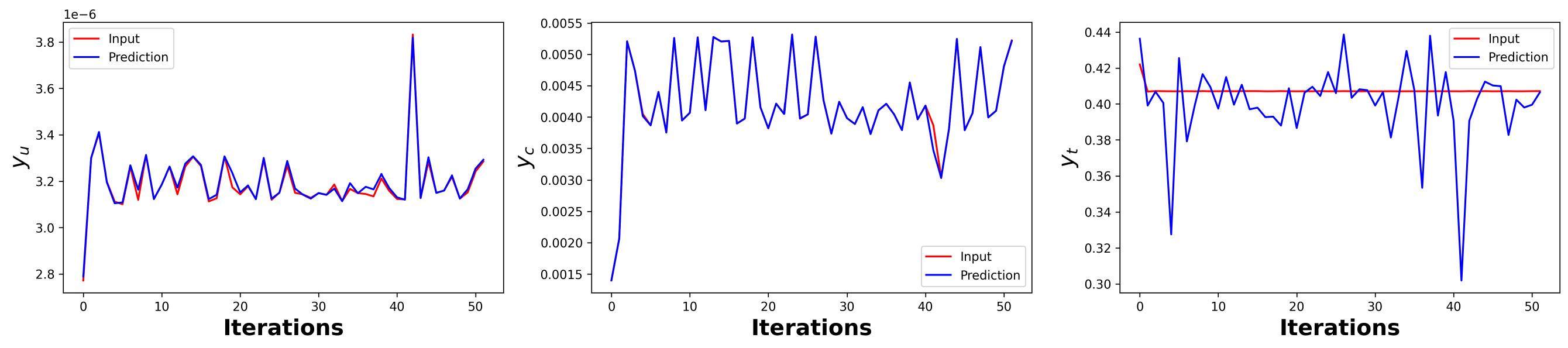}\\
    \includegraphics[width=1\linewidth]{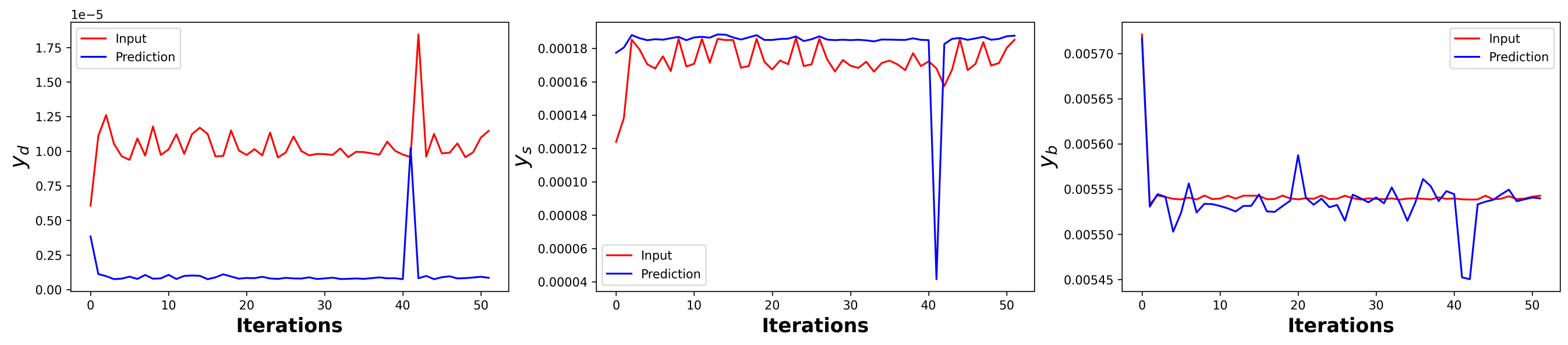}\\
\includegraphics[width=1\linewidth]{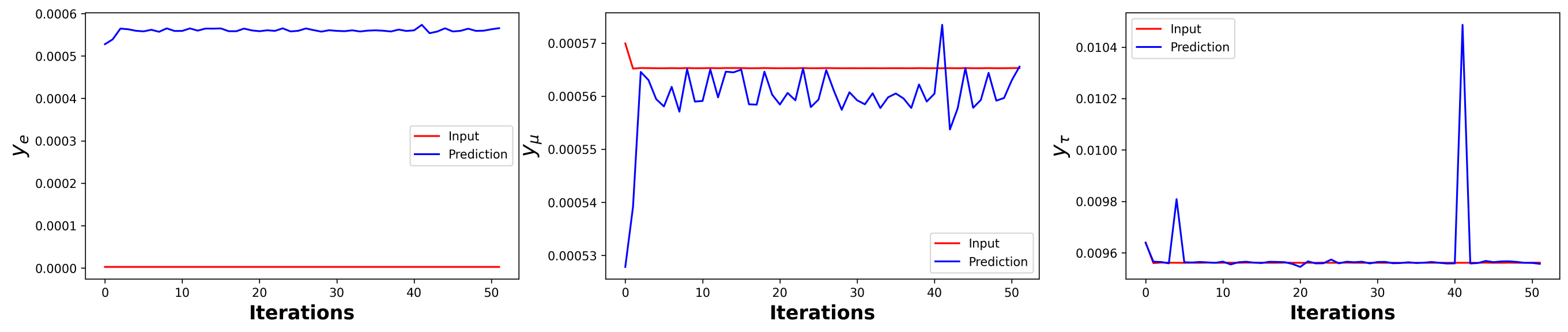}\\
\includegraphics[width=0.5\linewidth]{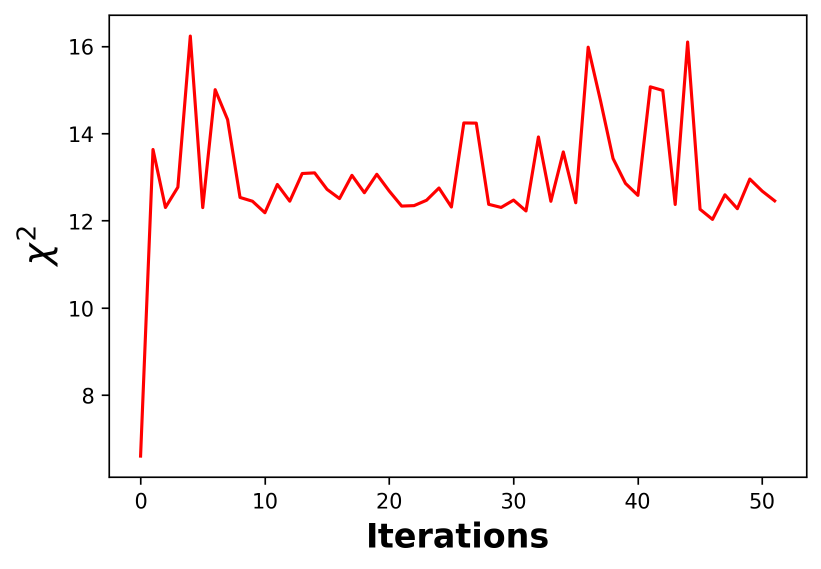}
\caption{The figure compares the input Yukawa couplings (red) with the predicted values (blue) at the matching scale across multiple iterations. The figure in the last line shows the obtained $\chi^2$ at different iteration, where the zeroth iteration is without considering the scalar contribution. }
    \label{fig:iterative:GUT}
\end{figure}
\begin{figure}[t!]
    \centering
    \includegraphics[width=1\linewidth]{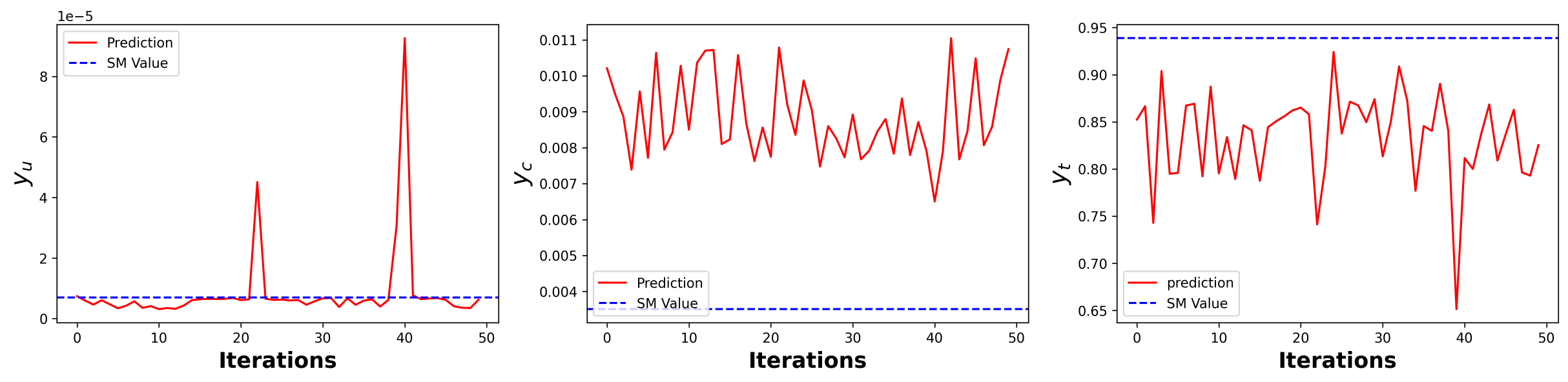}\\
        \includegraphics[width=1\linewidth]{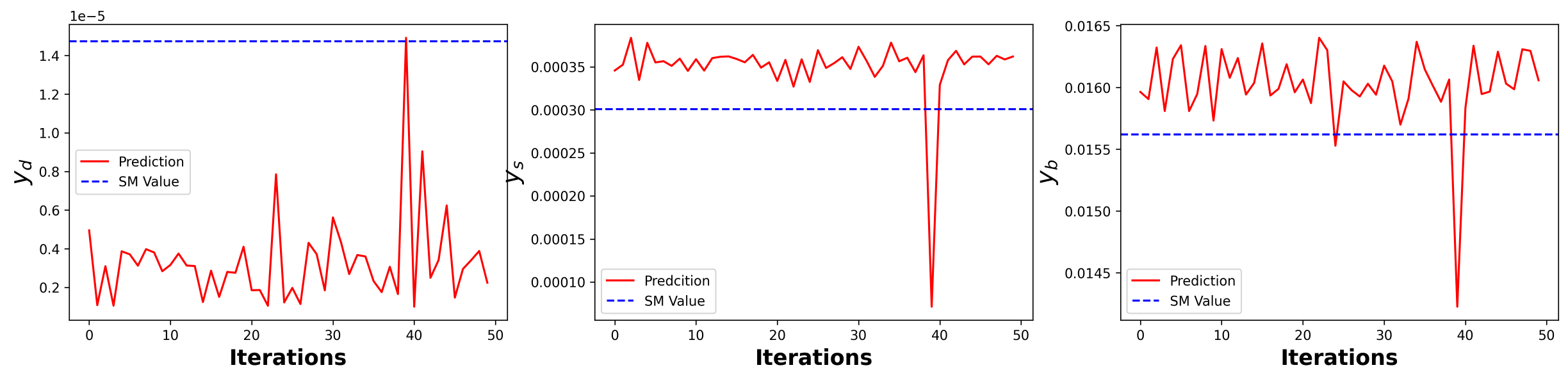}\\
            \includegraphics[width=1\linewidth]{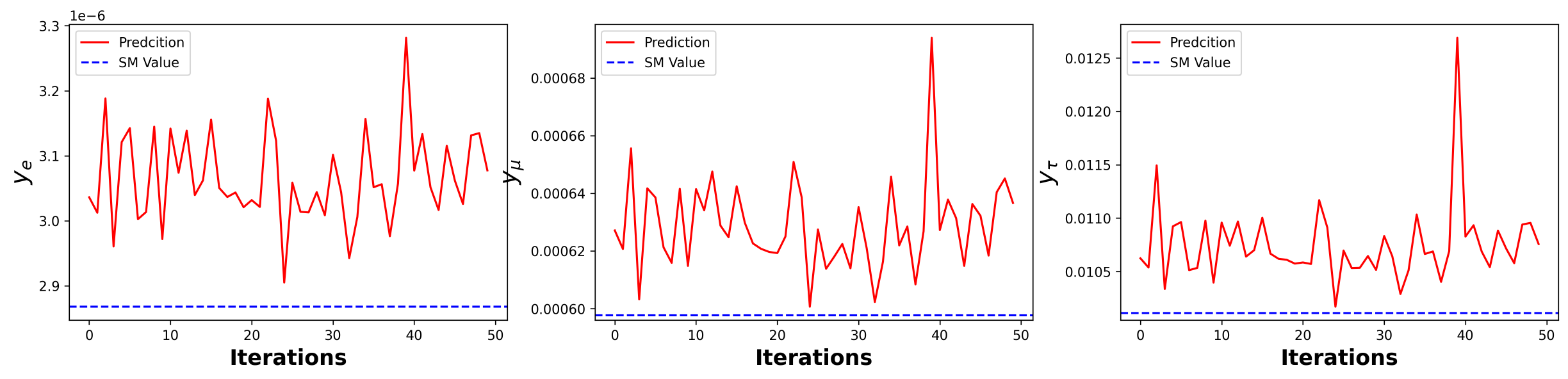}\\
                \includegraphics[width=1\linewidth]{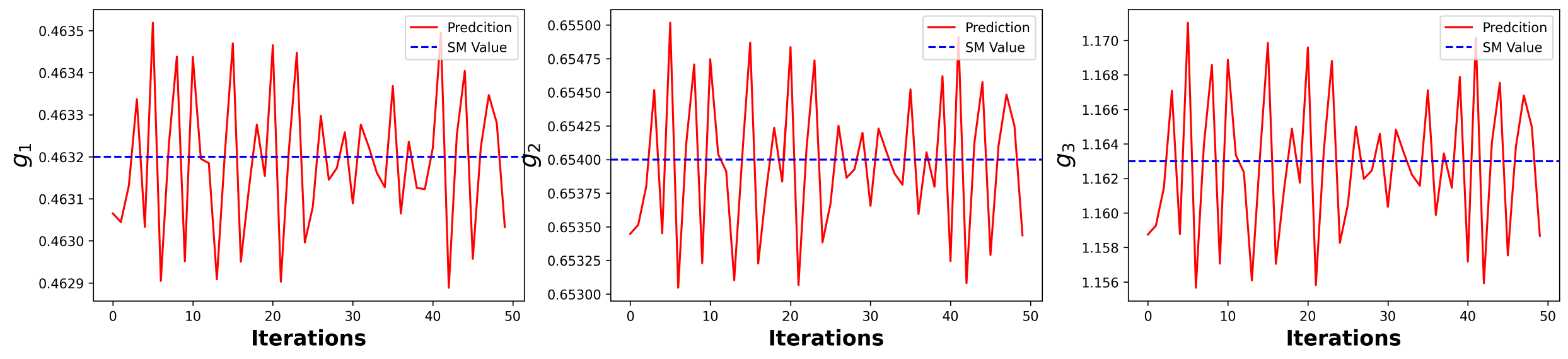}\\
                    \includegraphics[width=1\linewidth]{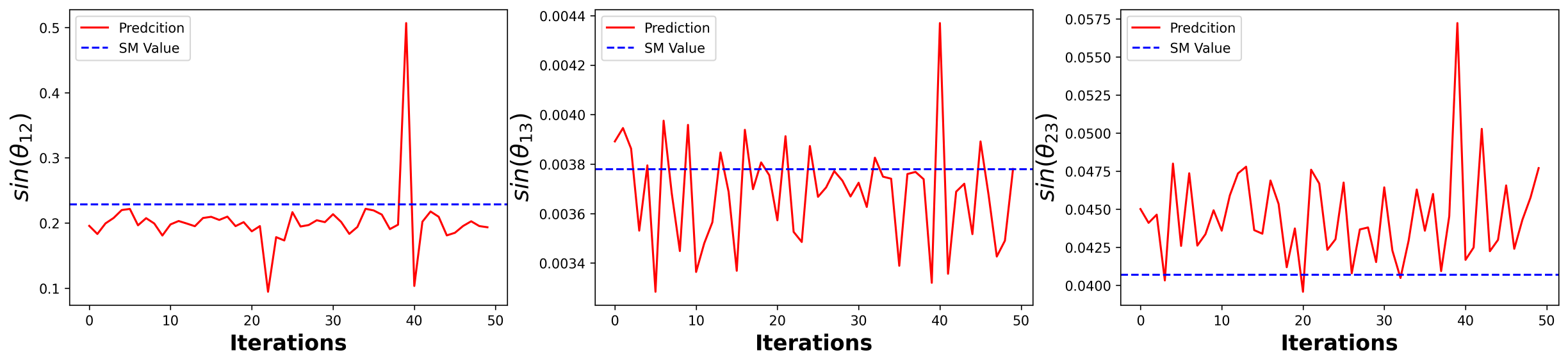}
   \caption{The figure compares the predicted Standard Model (SM) parameters (solid red lines) at the top quark mass scale ($M_t$) with the input SM parameter (blue dashed line).}
   \label{fig:iterative:SM}
\end{figure}

It has been noted above that some of the scalars deviate significantly from the matching scale~$(\mu)$ in order to viably reproduce the SM spectra. The mass of these scalars can, in principle, affect the renormalisation group evolution of the SM parameters and change their GUT scale values.  Below, the model-I has is considered (i.e Yukawa sector with only $45_\hh$) as it has least number of free parameters. It is evident from Tab.~(\ref{tab:sol1}), that the lightest scalar is $(O\big(8,2, \sfrac{1}{2}))$ and  contribution of the lightest scalar to the SM RGE is considered at its mass $M_O$, and the process is elaborated below:

\begin{enumerate}[label=\textbf{Step} \arabic*:]
    \item The SM parameters are extrapolated using two-loop RGE equations from $M_t$ (top pole mass) to the GUT scale and used for fitting the GUT scale parameters. This step (i.e., zeroth iteration) will yield the Yukawa matrices $Y_{1,2}$ and the masses of the various scalars listed in Tab.~(\ref{tab:sol1}). From here, the mass of the lightest scalar is determined, which is $M_O \sim 10^{7}$GeV.  
    \item The interaction of the scalar field $O$ with SM fermions is constructed as shown below; 
    \begin{equation}
        {\cal L} \supset Y_{O_1}\,q^T\,\cc\,u^C\,O + Y_{O_2}\,q^T\,\cc\,d^C\,O^{\dagger} + \text{h.c.}
    \end{equation}
    Since $O$ ($O^{\dagger}$) stems from $45_\hh$ ($45_\hh^{\dagger}$), the largest entry of $Y_1$ becomes the (1,1) element of $Y_{O_1}$, and the largest element of $Y_2$ becomes the (2,2) element of $Y_{O_2}$, while all other elements of $Y_{O_{1,2}}$ are taken to be zero\footnote{This particular choice of $Y_{O_{1,2}}$ ensures no large contribution of the scalar field's Yukawa coupling to the SM parameters.}. It is to be noted that coupling of $Y_{O_{1,2}}$ is also complex in nature.
    \item Having constructed the $Y_{O_{1,2}}$ Yukawa couplings from $Y_{1,2}$, the parameters are run down from $M_{\rm{GUT}}$ to $M_O$ (determined in Step 1) using two-loop SM + Octet RGE equations. The running parameters include the fitted values of $Y_{u,d,e}$, the formulated values of $Y_{O_{1,2}}$, and the computed values of $g_{1,2,3}$\footnote{The considered scenario does not achieve gauge coupling unification. Therefore, while performing the fit, $g$ is fixed at $g \equiv g_5 = 0.524$, but for the beta function, the values of $g_{1,2,3}$ at the respective scale obtained from running are used.}. 
    \item After obtaining the values of $Y_{O_{1,2}}$ at $M_{O}$, these values are inserted as initial conditions for the two-loop SM + Octet RGE equations and solved from $M_O$ to $\mu = M_{\rm{GUT}}$. The initial conditions for these RGE equations include the solutions for $Y_{u,d,e}$ and $g_{1,2,3}$ obtained from solving the SM RGE equations from $M_t$ to $M_O$. Now, $Y_{u,d,e}$ and CKM parameters are obtained at the GUT scale, incorporating the contribution of the octet scalar.
    \item Since the octet contribution has been included in the RGE equations, its threshold correction at the matching scale has been set to zero. With this, the obtained SM parameters are fitted at the GUT scale, yielding a new set of $Y_{1,2}$ (first iteration). After this, Steps (2–5) are repeated for fifty iterations. While fitting the parameters all the Yukawa couplings are restricted to lie between $\big(-3.5,\,3.5\big)$. The Yukawa couplings $Y_{u,d,e}$ predicted at the GUT scale, when evolved down to the octet mass scale using SM + octet RGEs, are again run down to $M_t$ to yield SM parameters at $M_t$.  The resulting input values and the predicted SM parameters, both at the matching scale $(M_{\rm{GUT}})$ and $M_t$, are extracted and presented in Figs.~(\ref{fig:iterative:GUT} and \ref{fig:iterative:SM}).
\end{enumerate}

The above analysis demonstrates that the obtained solution can approximately reproduce the SM parameters, even after accounting for the contribution of the lightest scalar in the RGE evolution of the SM parameters. The top Yukawa coupling exhibits the largest deviation, both at the GUT scale and at $M_t$. A specific form of the Yukawa coupling of the scalar field $O$ is chosen to minimise its impact on the SM parameters. Any alternative choice would lead to different SM parameters at the GUT scale, and if these deviations are significant, their low-energy equivalents would differ substantially from the original values. Moreover, it is observed that the $\chi^2$ value increases when the octet's contribution to the RGEs is included. This occurs because its threshold correction contribution is set to zero at the GUT scale.

\subsection{Model I-A}
\label{ssec:solmodelII}
The viability of Eq.~(\ref{eq:SU5NM45n15}) in conjunction with Eq.~(\ref{eq:oneloopsu5}) is assessed in reproducing the observed charged and neutral fermion mass spectrum and mixing angles through \(\chi^2\) optimisation. For this purpose, the basis for \(Y_1\) and \(Y_2\) is chosen as before, with \(Y_3\) set as a real-symmetric matrix. The ability of Eqs.~(\ref{eq:SU5NM45n15}) and (\ref{eq:oneloopsu5}) to yield six quark Yukawas, three charged lepton Yukawas, four CKM parameters, two neutral fermion mass-squared differences, and four PMNS parameters is examined. The reference values of neutral lepton parameters are sourced from a recent update of $NuFIT-6.0$~\cite{Esteban:2024eli}.\\

 As done in the previous subsection~(\ref{ssec:solmodelI}), two-loop RGE-extrapolated values for charged fermion Yukawas are used for fitting. Although \(Y_u\) and \(Y_d\) appearing in Eq.~(\ref{eq:SU5NM45n15}) are at the GUT scale, the neutral lepton parameters are fitted at the low scale, assuming negligible running effects~\cite{Chankowski:1993tx,Babu:1993qv,Antusch:2005gp,Mei:2005qp}.  Choosing \( Y_3 \) as a real symmetric matrix adds six more Yukawa parameters, while \( 15_{\hh} \) introduces three additional scalars, a cubic coupling \((\lambda)\), and the \(vev\) of \( t \), all of which contribute additional new parameters.  This results in a total of 26 parameters to be fitted to reproduce 18 observables. Again all the Yukawa couplings are varied with the perturbative limit specified in the beginning of this section, masses of scalars are varied within $1$ TeV and the Plank scale, while mass of the gauge boson is equal to the matching scale~$(\mu\,=\,M_{GUT}\,=\,M_X\,=\,10^{16})$ GeV. The $\lambda$ and $\langle t \rangle $ parameter appearing in Eq.~(\ref{eq:SU5NM45n15}) are also varied near the GUT scale and $(10^{-2}-5)$ GeV respectively. Additionally, the vacuum expectation value of \(H\) is set at the electroweak scale~$( \rm{i.e.}~\langle H \rangle = 174\,\rm{GeV})$. The discussion of subsection~(\ref{sssec:protondecay}) puts lower bound on the mass of scalars $T$ and $\mathbb{T}$ which also has been taken care while fitting the parameters.  The best fitted of various parameters, in this case, are labelled as solution-IA in Tab.~(\ref{tab:sol1}).\\ 

The fitted values of $Y_{1,2,3}$ corresponding to the solution~(II) given in Tab.~(\ref{tab:sol1}) are as follows:
\beqa\label{eq:y1y2y345+15}
Y_1 \eq\begin{footnotesize} \begin{pmatrix}
    0 & \big(3.76776 + i\,0.968243\big)\times10^{-5} &  \big(1.95037 + i\,0.984594\big)\\
    \big(-3.76776 - i\,0.968243\big)\times10^{-4} & 0 & -1.09659 - i\,0.553524 \\
    \big(-1.95037 - i\,0.984594\big) & 1.09659 + i\,0.553524 & 0
    
\end{pmatrix}\,\end{footnotesize} ,\nl
Y_2 \eq \begin{pmatrix}
    -8.03887 \times 10^{-5} & 0 & 0\\
    0 & -2.51338 \times 10^{-3} & 0\\
    0 & 0 & -1.01186
\end{pmatrix}\,,\;\rm{and}, \nl
Y_3 \eq \begin{pmatrix}
    0.374622 & 0.302098 & -2.04874 \\
    0.302098  & 2.10474 & 0.377314 \\
    -2.04874  & 0.377314 & -0.318043
\end{pmatrix}\,.
\eeqa

 It is evident from the solution~I-A, given in Tab.~(\ref{tab:sol1}), that the entire charged and neutral fermion mass spectrum, along with the mixing angles, can be adequately fitted by including \(15_{\hh}\) in addition to \(45_{\hh}\) in the Yukawa sector. The entries of \(Y_3\) are of \({\cal O}(1)\) in Eq.~(\ref{eq:y1y2y345+15}) and are essential for reproducing the correct neutrino mass scales. The scalar fields $S$ deviates the most from the matching scale and of the $TeV$ order, while the other scalars are closer to the matching scale. The fitted value of the $vev$ of scalar field $t$ is close to $1.00$ GeV and the obtained $\chi^2$ corresponding to the fitted parameters is $2.0$.\\
 
 It is noteworthy that even in this case, the required mass of the scalar field \( {\cal T} \) is close to the GUT scale, complying with the any constraints that may arise from loop-induced proton decays. Furthermore, the scalar field \( \Omega \) is also near the GUT scale, ensuring that the resulting state from the mixing of \( \mathbb{T} \) and \( \Omega \) remains of the order of the GUT scale and satisfies proton decay constraints.

\subsection{Model II}

Finally, model-II given in subsection~(\ref{ssec:modelIII}) is assessed and the ability of Eqs.~(\ref{eq:SU15atoneloop}), (\ref{eq:wfrfrom15}), and (\ref{eq:NM15+5}) to reproduce the observed charged and neutral fermion mass spectra, as also performed in previous subsections, through \(\chi^2\) analysis. The chosen basis includes \(Y_1\) as a complex-symmetric matrix, \(Y_2\) as a real-diagonal matrix, and \(Y_3\) as a real-symmetric matrix, resulting in a total of 21 unknown Yukawa parameters to be fitted. The Yukawa couplings are varied within the range \((-1.0, 1.0)\), and the matching scale \((\mu)\) is fixed at the GUT scale, which is also set equal to the mass of the heavy gauge boson, consistent with previous settings. Additionally, the scalar masses are varied between 1 TeV and the Planck scale. The scalar field \(T\), known to induce tree-level proton decay, has its lower bound set to satisfy proton decay constraints~\cite{Dorsner:2012uz}. The \(\lambda\) parameter appearing in \(M_{\nu}\) of Eq.~(\ref{eq:NM15+5}) is varied again close to the GUT scale. The total number of unknown parameters to be fitted includes the Yukawa couplings, scalar masses, and \(\lambda\) parameter, giving a total of 26 parameters. The results of the \(\chi^2\) optimisation are shown in Tab.~(\ref{tab:sol2}) as the solution (II), and the fitted values of \(Y_{1,2,3}\) are provided in corresponding to the solution~(III) is as follows:

\beqa \label{eq:y1y2y35n15}
Y_1 \eq \begin{footnotesize} \begin{pmatrix}
    \big(-4.5031 - i\,5.5048\big)\times10^{-5} & \big(7.6529 - i\,3.8840\big)\times10^{-4} & \big(-3.0386 + i\,3.4412\big)\times 10^{-3}\\
    \big(7.6529 - i\,3.8840\big)\times10^{-4} &
    \big(4.1431 - i\,1.1378\big)\times10^{-3} & 
    \big(-3.2919 + i\,3.3610\big)\times10^{-2}\\
    \big(-3.0386 + i\,3.4412\big)\times 10^{-3}&
    \big(-3.2919 + i\,3.3610\big)\times10^{-2}&
    \big(1.7848 - i\,3.7993\big)\times10^{-1}
\end{pmatrix}\,,\end{footnotesize}\nl
Y_2 \eq \begin{pmatrix}
    -1.3211 \times 10^{-5} & 0 & 0\\
    0 & 1.2873 \times 10^{-2} & 0 \\
    0 & 0 & 1.1107\times 10^{-1}
\end{pmatrix}\,,\rm{and}\nl
Y_3 \eq \begin{pmatrix}
-0.21650 & -0.99153 & 0.71898 \\
-0.99153 & -0.31605 & -0.71097\\
0.71898 & -0.71097 & -0.76757
    
\end{pmatrix}\,.
\eeqa

\begin{table*}[t]
    \begin{center} 
        \begin{math} 
            \begin{tabular}{cccr}
                \hline
                \hline 
                \multirow{2}{*}{~~Observable~~} & \multirow{2}{*}{ \hspace*{0.5cm}  $O_{\rm exp}$ \hspace*{0.5cm} } &
                \multicolumn{1}{c}{\bf \hspace*{0.5cm} Solution II \hspace*{0.5cm} } \\
                 &   & $O_{\rm th}$  \\
                \hline
                $y_u$  & $2.81 \times 10^{-6}$ & $2.81\times 10^{-6}$  \\
                $y_c$   & $1.42\times10^{-3}$& $1.42\times 10^{-3}$\\
                $y_t$   & $4.27 \times 10^{-1} $ & $4.27 \times 10^{-1}$\\
                $y_d$  & $ 6.14 \times 10^{-6} $ & $6.12\times10^{-6}$  \\
                $y_s$   & $1.25 \times10^{-4}$ & $1.23\times10^{-4}$  \\
                $y_b$   & $5.80 \times10^{-3}$ & $5.83 \times10^{-3}$\\
                $y_e$   & $2.75 \times 10^{-6}$ & $2.75 \times 10^{-6}$  \\
                $y_{\mu}$  &$5.72 \times10^{-4}$ & $5.73\times10^{-4}$ \\
                $y_{\tau}$   & $9.68 \times10^{-3}$ & $9.62 \times10^{-3}$\\
                $|V_{us}|$ & $0.2286$ & $0.2288$  \\
                $|V_{cb}|$ & $0.0457$ & $0.0458$ \\
                $|V_{ub}|$ & $0.0042$ & $0.0042$  \\
                $\sin\delta_{\rm CKM}$ & $0.78$ & $0.78$  \\
                $\Delta m^2_{\text{sol}}~ [{\rm eV}^2]$ & $7.49\times10^{-5}$& $7.84 \times 10^{-5}$  \\
				$\Delta m^2_{\text{atm}}~ [{\rm eV}^2]$ & $2.534\times10^{-3}$& $2.65 \times 10^{-3}$   \\
				$\sin^2 \theta _{12}$ & $0.307$ & $0.306$  \\
				$\sin^2 \theta _{23}$ & $0.561$ & $0.559$  \\
				$\sin^2 \theta _{13}$ & $0.02195$ & $0.02195$ \\

                \hline
                $\chi^2_{\rm min}$    & &  $\sim 0$ \\
                \hline
                $M_T$ [GeV]& &$2.34\times10^{14}$ & \\
                $M_{\Delta}$ [GeV]& &$1.01\times 10^{3}$ & \\
                $M_{S}$ [GeV]& &$3.26\times 10^{8}$ & \\   
                $M_{t}$ [GeV]& &$1.82\times 10^{15}$ & \\
                $\lambda$ [GeV]& &$7.98\times 10^{15}$ & \\
                \hline
                \hline 
            \end{tabular}
        \end{math}
    \end{center}
    \caption{The benchmark best-fit values of different observables for the charged and neutral fermion mass spectrum and quark and lepton mixing parameters. Other details are same as in Tab.~(\ref{tab:sol1}).} 
    \label{tab:sol2} 
\end{table*}

The Yukawa entries \(Y_{1,2,3}\) corresponding to solution~(II) are provided in Eq.~(\ref{eq:y1y2y35n15}). As there are 26 unknown parameters, the entries of \(Y_{1,2,3}\) are feeble and well within the perturbative range. Among the scalars, \(M_{\Delta}\) shows the greatest deviation from the matching scale, while other scalars are closer to the matching scale. The fitted value of the trilinear coupling is approximately \(10^{16}\) GeV, and the corresponding \(\chi^2\) for the fitted values is \(\sim 0\).

\section{Summary}
\label{sec:summary}
In general, tree-level conclusions in any model—particularly Yukawa relations—can sometimes be unreliable and may not fully represent the actual situation, as also demonstrated by this article. Tree-level statements alone cannot capture the full dynamics of any model and must be augmented with loop corrections to provide a more accurate picture. The Yukawa sector of a renormalisable \(SU(5)\) GUT model has been comprehensively analysed in this study across different minimal settings, with a focus on the inclusion of quantum corrections. The quantum corrections to the various Yukawa vertices are induced by heavy degrees of freedom—scalar fields and heavy gauge bosons. These heavy degrees of freedom impart threshold corrections at a particular matching scale. The noteworthy and revealing features of this study are as follows:
\begin{itemize}
\item The Yukawa sector of an \(SU(5)\) model containing only the \(45_{\hh}\)-dimensional irrep can reproduce the correct charged lepton spectrum when one-loop corrections to various Yukawa vertices are included. Furthermore, adding the \(15_{\hh}\) irrep to the Yukawa sector alongside the \(45_{\hh}\) yields the observed charged and neutral fermion mass spectra and mixing angles when quantum corrections are switched on, even after considering the RGE effects of lightest scalars.
  
\item Incorporating loop corrections to the Yukawa relations of an $SU(5)$ model comprising of \(5_{\hh}\) and \(15_{\hh}\) dimensional irreps also successfully reproduces the correct fermion mass spectrum and mixing angles.\\
    
\end{itemize}
A general feature of this analysis is that, to achieve a desirable Yukawa spectrum, either larger Yukawa couplings are needed or the mass of one or more scalar particles must significantly deviate from the matching scale, as these are the primary sources for achieving the necessary threshold corrections. An appealing aspect of GUTs is that they possess multiple heavy degrees of freedom, often arising from a single irrep, which can contribute to Yukawa relations at loop level, as harnessed in this study. However, for a viable spectrum, these heavy degrees of freedom—i.e., scalar fields—require mass splitting, with many needing significant deviation from the matching scale. This scenario calls for a mechanism to generate mass splittings in non-supersymmetric GUTs at all loop levels. \\

As the focus of this study is to construct a minimal Yukawa sector, other important aspects like gauge coupling unification have not been addressed. To achieve gauge coupling unification, one can include additional (scalar and fermion) irreps that modify the RGE of gauge couplings without affecting the Yukawa sector~\cite{Cox:2016epl}. It is noteworthy that quantum corrections can significantly alter the tree-level Yukawa relations, making them phenomenologically desirable. This serves as a motivation to explore the Yukawa sectors of other GUT models.



\section*{Acknowledgment}
 I express my sincere gratitude to Ketan M. Patel for discussions, suggestions and review of the manuscript. Additionally, I also acknowledge discussions with Debashis Pachhar. I would also like to thank (different) anonymous referee(s) for the valuable comments, which have significantly improved this manuscript.  This work was carried out at the Physical Research Laboratory, Gujarat, India, with support from the Department of Space (DOS), Government of India.

 \section*{Appendix}
\appendix

\section{Loop integration factors}
\label{app:LF}
The loop integration factors appearing in Eqs. (\ref{eq:deltaf}, \ref{eq:kfs}, \ref{eq:wfrfrom15}, \ref{eq:SU5NM45n15}, \ref{eq:dY}, \ref{eq:K}) and (\ref{eq:NM15+5}) are given by;
\beqa \label{eq:LF_f}
f[m_1^2,m_2^2] &=& -\frac{1}{16 \pi^2} \left(\frac{m_1^2 \log \frac{m_1^2}{\mu^2}-m_2^2\log \frac{m_2^2}{\mu^2}}{m_1^2-m_2^2} - 1 \right)\,, \eeqa
\beqa \label{eq:LF_h}
h[m_1^2,m_2^2] &=& \frac{1}{16 \pi^2} \left(\frac{1}{2} \log\frac{m_1^2}{\mu^2} +  \frac{\frac{1}{2} r^2 \log r -\frac{3}{4}r^2 + r - \frac{1}{4}}{(1-r)^2}\right)\,, \eeqa

\beqa \label{eq:LF_g}
g[m_1^2,m_2^2] & = & \frac{1}{16 \pi^2} \frac{\frac{r^3}{6}-r^2+\frac{r}{2}+ r \log r + \frac{1}{3}}{(1 -r)^3}\,,\eeqa
where $r=m_2^2/m_1^2$ in the last two equations, and,
\beqa{\label{eq:LF_i}}
p\big(m_1^2,m_2^2\big) \eq \frac{1}{16\pi^2}\,\frac{1}{m_1^2 - m_2^2} \log\left(\frac{m_1^2}{m_2^2}\right)
\eeqa


\bibliography{references}

\begin{thebibliography}{67}%
\makeatletter
\providecommand \@ifxundefined [1]{%
 \@ifx{#1\undefined}
}%
\providecommand \@ifnum [1]{%
 \ifnum #1\expandafter \@firstoftwo
 \else \expandafter \@secondoftwo
 \fi
}%
\providecommand \@ifx [1]{%
 \ifx #1\expandafter \@firstoftwo
 \else \expandafter \@secondoftwo
 \fi
}%
\providecommand \natexlab [1]{#1}%
\providecommand \enquote  [1]{``#1''}%
\providecommand \bibnamefont  [1]{#1}%
\providecommand \bibfnamefont [1]{#1}%
\providecommand \citenamefont [1]{#1}%
\providecommand \href@noop [0]{\@secondoftwo}%
\providecommand \href [0]{\begingroup \@sanitize@url \@href}%
\providecommand \@href[1]{\@@startlink{#1}\@@href}%
\providecommand \@@href[1]{\endgroup#1\@@endlink}%
\providecommand \@sanitize@url [0]{\catcode `\\12\catcode `\$12\catcode `\&12\catcode `\#12\catcode `\^12\catcode `\_12\catcode `\%12\relax}%
\providecommand \@@startlink[1]{}%
\providecommand \@@endlink[0]{}%
\providecommand \url  [0]{\begingroup\@sanitize@url \@url }%
\providecommand \@url [1]{\endgroup\@href {#1}{\urlprefix }}%
\providecommand \urlprefix  [0]{URL }%
\providecommand \Eprint [0]{\href }%
\providecommand \doibase [0]{http://dx.doi.org/}%
\providecommand \selectlanguage [0]{\@gobble}%
\providecommand \bibinfo  [0]{\@secondoftwo}%
\providecommand \bibfield  [0]{\@secondoftwo}%
\providecommand \translation [1]{[#1]}%
\providecommand \BibitemOpen [0]{}%
\providecommand \bibitemStop [0]{}%
\providecommand \bibitemNoStop [0]{.\EOS\space}%
\providecommand \EOS [0]{\spacefactor3000\relax}%
\providecommand \BibitemShut  [1]{\csname bibitem#1\endcsname}%
\let\auto@bib@innerbib\@empty
\bibitem [{\citenamefont {Georgi}\ and\ \citenamefont {Glashow}(1974)}]{Georgi:1974sy}%
  \BibitemOpen
  \bibfield  {author} {\bibinfo {author} {\bibfnamefont {H.}~\bibnamefont {Georgi}}\ and\ \bibinfo {author} {\bibfnamefont {S.~L.}\ \bibnamefont {Glashow}},\ }\bibfield  {title} {\enquote {\bibinfo {title} {{Unity of All Elementary Particle Forces}},}\ }\href {\doibase 10.1103/PhysRevLett.32.438} {\bibfield  {journal} {\bibinfo  {journal} {Phys. Rev. Lett.}\ }\textbf {\bibinfo {volume} {32}},\ \bibinfo {pages} {438--441} (\bibinfo {year} {1974})}\BibitemShut {NoStop}%
\bibitem [{\citenamefont {Fritzsch}\ and\ \citenamefont {Minkowski}(1975)}]{Fritzsch:1974nn}%
  \BibitemOpen
  \bibfield  {author} {\bibinfo {author} {\bibfnamefont {Harald}\ \bibnamefont {Fritzsch}}\ and\ \bibinfo {author} {\bibfnamefont {Peter}\ \bibnamefont {Minkowski}},\ }\bibfield  {title} {\enquote {\bibinfo {title} {{Unified Interactions of Leptons and Hadrons}},}\ }\href {\doibase 10.1016/0003-4916(75)90211-0} {\bibfield  {journal} {\bibinfo  {journal} {Annals Phys.}\ }\textbf {\bibinfo {volume} {93}},\ \bibinfo {pages} {193--266} (\bibinfo {year} {1975})}\BibitemShut {NoStop}%
\bibitem [{\citenamefont {Gell-Mann}\ \emph {et~al.}(1979)\citenamefont {Gell-Mann}, \citenamefont {Ramond},\ and\ \citenamefont {Slansky}}]{Gell-Mann:1979vob}%
  \BibitemOpen
  \bibfield  {author} {\bibinfo {author} {\bibfnamefont {Murray}\ \bibnamefont {Gell-Mann}}, \bibinfo {author} {\bibfnamefont {Pierre}\ \bibnamefont {Ramond}}, \ and\ \bibinfo {author} {\bibfnamefont {Richard}\ \bibnamefont {Slansky}},\ }\bibfield  {title} {\enquote {\bibinfo {title} {{Complex Spinors and Unified Theories}},}\ }\href@noop {} {\bibfield  {journal} {\bibinfo  {journal} {Conf. Proc. C}\ }\textbf {\bibinfo {volume} {790927}},\ \bibinfo {pages} {315--321} (\bibinfo {year} {1979})},\ \Eprint {http://arxiv.org/abs/1306.4669} {arXiv:1306.4669 [hep-th]} \BibitemShut {NoStop}%
\bibitem [{\citenamefont {Langacker}(1981)}]{Langacker:1980js}%
  \BibitemOpen
  \bibfield  {author} {\bibinfo {author} {\bibfnamefont {Paul}\ \bibnamefont {Langacker}},\ }\bibfield  {title} {\enquote {\bibinfo {title} {{Grand Unified Theories and Proton Decay}},}\ }\href {\doibase 10.1016/0370-1573(81)90059-4} {\bibfield  {journal} {\bibinfo  {journal} {Phys. Rept.}\ }\textbf {\bibinfo {volume} {72}},\ \bibinfo {pages} {185} (\bibinfo {year} {1981})}\BibitemShut {NoStop}%
\bibitem [{\citenamefont {Nath}\ and\ \citenamefont {Fileviez~Perez}(2007)}]{Nath:2006ut}%
  \BibitemOpen
  \bibfield  {author} {\bibinfo {author} {\bibfnamefont {Pran}\ \bibnamefont {Nath}}\ and\ \bibinfo {author} {\bibfnamefont {Pavel}\ \bibnamefont {Fileviez~Perez}},\ }\bibfield  {title} {\enquote {\bibinfo {title} {{Proton stability in grand unified theories, in strings and in branes}},}\ }\href {\doibase 10.1016/j.physrep.2007.02.010} {\bibfield  {journal} {\bibinfo  {journal} {Phys. Rept.}\ }\textbf {\bibinfo {volume} {441}},\ \bibinfo {pages} {191--317} (\bibinfo {year} {2007})},\ \Eprint {http://arxiv.org/abs/hep-ph/0601023} {arXiv:hep-ph/0601023} \BibitemShut {NoStop}%
\bibitem [{\citenamefont {Georgi}\ and\ \citenamefont {Jarlskog}(1979)}]{Georgi:1979df}%
  \BibitemOpen
  \bibfield  {author} {\bibinfo {author} {\bibfnamefont {Howard}\ \bibnamefont {Georgi}}\ and\ \bibinfo {author} {\bibfnamefont {C.}~\bibnamefont {Jarlskog}},\ }\bibfield  {title} {\enquote {\bibinfo {title} {{A New Lepton - Quark Mass Relation in a Unified Theory}},}\ }\href {\doibase 10.1016/0370-2693(79)90842-6} {\bibfield  {journal} {\bibinfo  {journal} {Phys. Lett. B}\ }\textbf {\bibinfo {volume} {86}},\ \bibinfo {pages} {297--300} (\bibinfo {year} {1979})}\BibitemShut {NoStop}%
\bibitem [{\citenamefont {Barbieri}\ \emph {et~al.}(1981)\citenamefont {Barbieri}, \citenamefont {Nanopoulos},\ and\ \citenamefont {Wyler}}]{Barbieri:1981yw}%
  \BibitemOpen
  \bibfield  {author} {\bibinfo {author} {\bibfnamefont {Riccardo}\ \bibnamefont {Barbieri}}, \bibinfo {author} {\bibfnamefont {Dimitri~V.}\ \bibnamefont {Nanopoulos}}, \ and\ \bibinfo {author} {\bibfnamefont {D.}~\bibnamefont {Wyler}},\ }\bibfield  {title} {\enquote {\bibinfo {title} {{Hierarchical Fermion Masses in SU(5)}},}\ }\href {\doibase 10.1016/0370-2693(81)90076-9} {\bibfield  {journal} {\bibinfo  {journal} {Phys. Lett. B}\ }\textbf {\bibinfo {volume} {103}},\ \bibinfo {pages} {433--436} (\bibinfo {year} {1981})}\BibitemShut {NoStop}%
\bibitem [{\citenamefont {Giveon}\ \emph {et~al.}(1991)\citenamefont {Giveon}, \citenamefont {Hall},\ and\ \citenamefont {Sarid}}]{Giveon:1991zm}%
  \BibitemOpen
  \bibfield  {author} {\bibinfo {author} {\bibfnamefont {Amit}\ \bibnamefont {Giveon}}, \bibinfo {author} {\bibfnamefont {Lawrence~J.}\ \bibnamefont {Hall}}, \ and\ \bibinfo {author} {\bibfnamefont {Uri}\ \bibnamefont {Sarid}},\ }\bibfield  {title} {\enquote {\bibinfo {title} {{SU (5) unification revisited}},}\ }\href {\doibase 10.1016/0370-2693(91)91289-8} {\bibfield  {journal} {\bibinfo  {journal} {Phys. Lett. B}\ }\textbf {\bibinfo {volume} {271}},\ \bibinfo {pages} {138--144} (\bibinfo {year} {1991})}\BibitemShut {NoStop}%
\bibitem [{\citenamefont {Dorsner}\ and\ \citenamefont {Fileviez~Perez}(2006)}]{Dorsner:2006dj}%
  \BibitemOpen
  \bibfield  {author} {\bibinfo {author} {\bibfnamefont {Ilja}\ \bibnamefont {Dorsner}}\ and\ \bibinfo {author} {\bibfnamefont {Pavel}\ \bibnamefont {Fileviez~Perez}},\ }\bibfield  {title} {\enquote {\bibinfo {title} {{Unification versus proton decay in SU(5)}},}\ }\href {\doibase 10.1016/j.physletb.2006.09.034} {\bibfield  {journal} {\bibinfo  {journal} {Phys. Lett. B}\ }\textbf {\bibinfo {volume} {642}},\ \bibinfo {pages} {248--252} (\bibinfo {year} {2006})},\ \Eprint {http://arxiv.org/abs/hep-ph/0606062} {arXiv:hep-ph/0606062} \BibitemShut {NoStop}%
\bibitem [{\citenamefont {Fileviez~Perez}(2007)}]{FileviezPerez:2007bcw}%
  \BibitemOpen
  \bibfield  {author} {\bibinfo {author} {\bibfnamefont {Pavel}\ \bibnamefont {Fileviez~Perez}},\ }\bibfield  {title} {\enquote {\bibinfo {title} {{Renormalizable adjoint SU(5)}},}\ }\href {\doibase 10.1016/j.physletb.2007.07.075} {\bibfield  {journal} {\bibinfo  {journal} {Phys. Lett. B}\ }\textbf {\bibinfo {volume} {654}},\ \bibinfo {pages} {189--193} (\bibinfo {year} {2007})},\ \Eprint {http://arxiv.org/abs/hep-ph/0702287} {arXiv:hep-ph/0702287} \BibitemShut {NoStop}%
\bibitem [{\citenamefont {Goto}\ \emph {et~al.}(2023)\citenamefont {Goto}, \citenamefont {Mishima},\ and\ \citenamefont {Shindou}}]{Goto:2023qch}%
  \BibitemOpen
  \bibfield  {author} {\bibinfo {author} {\bibfnamefont {Toru}\ \bibnamefont {Goto}}, \bibinfo {author} {\bibfnamefont {Satoshi}\ \bibnamefont {Mishima}}, \ and\ \bibinfo {author} {\bibfnamefont {Tetsuo}\ \bibnamefont {Shindou}},\ }\bibfield  {title} {\enquote {\bibinfo {title} {{Flavor physics in SU(5) GUT with a 45 scalar representation}},}\ }\href@noop {} {\  (\bibinfo {year} {2023})},\ \Eprint {http://arxiv.org/abs/2308.13329} {arXiv:2308.13329 [hep-ph]} \BibitemShut {NoStop}%
\bibitem [{\citenamefont {Ellis}\ and\ \citenamefont {Gaillard}(1979)}]{Ellis:1979fg}%
  \BibitemOpen
  \bibfield  {author} {\bibinfo {author} {\bibfnamefont {John~R.}\ \bibnamefont {Ellis}}\ and\ \bibinfo {author} {\bibfnamefont {Mary~K.}\ \bibnamefont {Gaillard}},\ }\bibfield  {title} {\enquote {\bibinfo {title} {{Fermion Masses and Higgs Representations in SU(5)}},}\ }\href {\doibase 10.1016/0370-2693(79)90476-3} {\bibfield  {journal} {\bibinfo  {journal} {Phys. Lett. B}\ }\textbf {\bibinfo {volume} {88}},\ \bibinfo {pages} {315--319} (\bibinfo {year} {1979})}\BibitemShut {NoStop}%
\bibitem [{\citenamefont {Berezinsky}\ and\ \citenamefont {Smirnov}(1984)}]{Berezinsky:1983va}%
  \BibitemOpen
  \bibfield  {author} {\bibinfo {author} {\bibfnamefont {V.~S.}\ \bibnamefont {Berezinsky}}\ and\ \bibinfo {author} {\bibfnamefont {A.~Yu.}\ \bibnamefont {Smirnov}},\ }\bibfield  {title} {\enquote {\bibinfo {title} {{HOW TO SAVE MINIMAL SU(5)}},}\ }\href {\doibase 10.1016/0370-2693(84)91044-X} {\bibfield  {journal} {\bibinfo  {journal} {Phys. Lett. B}\ }\textbf {\bibinfo {volume} {140}},\ \bibinfo {pages} {49--52} (\bibinfo {year} {1984})}\BibitemShut {NoStop}%
\bibitem [{\citenamefont {Altarelli}\ \emph {et~al.}(2000)\citenamefont {Altarelli}, \citenamefont {Feruglio},\ and\ \citenamefont {Masina}}]{Altarelli:2000fu}%
  \BibitemOpen
  \bibfield  {author} {\bibinfo {author} {\bibfnamefont {Guido}\ \bibnamefont {Altarelli}}, \bibinfo {author} {\bibfnamefont {Ferruccio}\ \bibnamefont {Feruglio}}, \ and\ \bibinfo {author} {\bibfnamefont {Isabella}\ \bibnamefont {Masina}},\ }\bibfield  {title} {\enquote {\bibinfo {title} {{From minimal to realistic supersymmetric SU(5) grand unification}},}\ }\href {\doibase 10.1088/1126-6708/2000/11/040} {\bibfield  {journal} {\bibinfo  {journal} {JHEP}\ }\textbf {\bibinfo {volume} {11}},\ \bibinfo {pages} {040} (\bibinfo {year} {2000})},\ \Eprint {http://arxiv.org/abs/hep-ph/0007254} {arXiv:hep-ph/0007254} \BibitemShut {NoStop}%
\bibitem [{\citenamefont {Emmanuel-Costa}\ and\ \citenamefont {Wiesenfeldt}(2003)}]{Emmanuel-Costa:2003szk}%
  \BibitemOpen
  \bibfield  {author} {\bibinfo {author} {\bibfnamefont {D.}~\bibnamefont {Emmanuel-Costa}}\ and\ \bibinfo {author} {\bibfnamefont {S.}~\bibnamefont {Wiesenfeldt}},\ }\bibfield  {title} {\enquote {\bibinfo {title} {{Proton decay in a consistent supersymmetric SU(5) GUT model}},}\ }\href {\doibase 10.1016/S0550-3213(03)00301-8} {\bibfield  {journal} {\bibinfo  {journal} {Nucl. Phys. B}\ }\textbf {\bibinfo {volume} {661}},\ \bibinfo {pages} {62--82} (\bibinfo {year} {2003})},\ \Eprint {http://arxiv.org/abs/hep-ph/0302272} {arXiv:hep-ph/0302272} \BibitemShut {NoStop}%
\bibitem [{\citenamefont {Dorsner}\ \emph {et~al.}(2007)\citenamefont {Dorsner}, \citenamefont {Fileviez~Perez},\ and\ \citenamefont {Rodrigo}}]{Dorsner:2006hw}%
  \BibitemOpen
  \bibfield  {author} {\bibinfo {author} {\bibfnamefont {Ilja}\ \bibnamefont {Dorsner}}, \bibinfo {author} {\bibfnamefont {Pavel}\ \bibnamefont {Fileviez~Perez}}, \ and\ \bibinfo {author} {\bibfnamefont {German}\ \bibnamefont {Rodrigo}},\ }\bibfield  {title} {\enquote {\bibinfo {title} {{Fermion masses and the UV cutoff of the minimal realistic SU(5)}},}\ }\href {\doibase 10.1103/PhysRevD.75.125007} {\bibfield  {journal} {\bibinfo  {journal} {Phys. Rev. D}\ }\textbf {\bibinfo {volume} {75}},\ \bibinfo {pages} {125007} (\bibinfo {year} {2007})},\ \Eprint {http://arxiv.org/abs/hep-ph/0607208} {arXiv:hep-ph/0607208} \BibitemShut {NoStop}%
\bibitem [{\citenamefont {Antusch}\ and\ \citenamefont {Hinze}(2022)}]{Antusch:2021yqe}%
  \BibitemOpen
  \bibfield  {author} {\bibinfo {author} {\bibfnamefont {Stefan}\ \bibnamefont {Antusch}}\ and\ \bibinfo {author} {\bibfnamefont {Kevin}\ \bibnamefont {Hinze}},\ }\bibfield  {title} {\enquote {\bibinfo {title} {{Nucleon decay in a minimal non-SUSY GUT with predicted quark-lepton Yukawa ratios}},}\ }\href {\doibase 10.1016/j.nuclphysb.2022.115719} {\bibfield  {journal} {\bibinfo  {journal} {Nucl. Phys. B}\ }\textbf {\bibinfo {volume} {976}},\ \bibinfo {pages} {115719} (\bibinfo {year} {2022})},\ \Eprint {http://arxiv.org/abs/2108.08080} {arXiv:2108.08080 [hep-ph]} \BibitemShut {NoStop}%
\bibitem [{\citenamefont {Antusch}\ \emph {et~al.}(2023{\natexlab{a}})\citenamefont {Antusch}, \citenamefont {Hinze},\ and\ \citenamefont {Saad}}]{Antusch:2022afk}%
  \BibitemOpen
  \bibfield  {author} {\bibinfo {author} {\bibfnamefont {Stefan}\ \bibnamefont {Antusch}}, \bibinfo {author} {\bibfnamefont {Kevin}\ \bibnamefont {Hinze}}, \ and\ \bibinfo {author} {\bibfnamefont {Shaikh}\ \bibnamefont {Saad}},\ }\bibfield  {title} {\enquote {\bibinfo {title} {{Viable quark-lepton Yukawa ratios and nucleon decay predictions in SU(5) GUTs with type-II seesaw}},}\ }\href {\doibase 10.1016/j.nuclphysb.2022.116049} {\bibfield  {journal} {\bibinfo  {journal} {Nucl. Phys. B}\ }\textbf {\bibinfo {volume} {986}},\ \bibinfo {pages} {116049} (\bibinfo {year} {2023}{\natexlab{a}})},\ \Eprint {http://arxiv.org/abs/2205.01120} {arXiv:2205.01120 [hep-ph]} \BibitemShut {NoStop}%
\bibitem [{\citenamefont {Hempfling}(1994)}]{Hempfling:1993kv}%
  \BibitemOpen
  \bibfield  {author} {\bibinfo {author} {\bibfnamefont {Ralf}\ \bibnamefont {Hempfling}},\ }\bibfield  {title} {\enquote {\bibinfo {title} {{Yukawa coupling unification with supersymmetric threshold corrections}},}\ }\href {\doibase 10.1103/PhysRevD.49.6168} {\bibfield  {journal} {\bibinfo  {journal} {Phys. Rev. D}\ }\textbf {\bibinfo {volume} {49}},\ \bibinfo {pages} {6168--6172} (\bibinfo {year} {1994})}\BibitemShut {NoStop}%
\bibitem [{\citenamefont {Shafi}\ and\ \citenamefont {Tavartkiladze}(1999)}]{Shafi:1999rm}%
  \BibitemOpen
  \bibfield  {author} {\bibinfo {author} {\bibfnamefont {Qaisar}\ \bibnamefont {Shafi}}\ and\ \bibinfo {author} {\bibfnamefont {Zurab}\ \bibnamefont {Tavartkiladze}},\ }\bibfield  {title} {\enquote {\bibinfo {title} {{Neutrino mixings and fermion masses in supersymmetric SU(5)}},}\ }\href {\doibase 10.1016/S0370-2693(99)00185-9} {\bibfield  {journal} {\bibinfo  {journal} {Phys. Lett. B}\ }\textbf {\bibinfo {volume} {451}},\ \bibinfo {pages} {129--135} (\bibinfo {year} {1999})},\ \Eprint {http://arxiv.org/abs/hep-ph/9901243} {arXiv:hep-ph/9901243} \BibitemShut {NoStop}%
\bibitem [{\citenamefont {Barr}\ and\ \citenamefont {Dorsner}(2003)}]{Barr:2003zx}%
  \BibitemOpen
  \bibfield  {author} {\bibinfo {author} {\bibfnamefont {S.~M.}\ \bibnamefont {Barr}}\ and\ \bibinfo {author} {\bibfnamefont {I.}~\bibnamefont {Dorsner}},\ }\bibfield  {title} {\enquote {\bibinfo {title} {{Explaining why the u and d quark masses are similar}},}\ }\href {\doibase 10.1016/S0370-2693(03)00772-X} {\bibfield  {journal} {\bibinfo  {journal} {Phys. Lett. B}\ }\textbf {\bibinfo {volume} {566}},\ \bibinfo {pages} {125--130} (\bibinfo {year} {2003})},\ \Eprint {http://arxiv.org/abs/hep-ph/0305090} {arXiv:hep-ph/0305090} \BibitemShut {NoStop}%
\bibitem [{\citenamefont {Oshimo}(2009)}]{Oshimo:2009ia}%
  \BibitemOpen
  \bibfield  {author} {\bibinfo {author} {\bibfnamefont {Noriyuki}\ \bibnamefont {Oshimo}},\ }\bibfield  {title} {\enquote {\bibinfo {title} {{Realistic model for SU(5) grand unification}},}\ }\href {\doibase 10.1103/PhysRevD.80.075011} {\bibfield  {journal} {\bibinfo  {journal} {Phys. Rev. D}\ }\textbf {\bibinfo {volume} {80}},\ \bibinfo {pages} {075011} (\bibinfo {year} {2009})},\ \Eprint {http://arxiv.org/abs/0907.3400} {arXiv:0907.3400 [hep-ph]} \BibitemShut {NoStop}%
\bibitem [{\citenamefont {Babu}\ \emph {et~al.}(2012)\citenamefont {Babu}, \citenamefont {Bajc},\ and\ \citenamefont {Tavartkiladze}}]{Babu:2012pb}%
  \BibitemOpen
  \bibfield  {author} {\bibinfo {author} {\bibfnamefont {K.~S.}\ \bibnamefont {Babu}}, \bibinfo {author} {\bibfnamefont {B.}~\bibnamefont {Bajc}}, \ and\ \bibinfo {author} {\bibfnamefont {Z.}~\bibnamefont {Tavartkiladze}},\ }\bibfield  {title} {\enquote {\bibinfo {title} {{Realistic Fermion Masses and Nucleon Decay Rates in SUSY SU(5) with Vector-Like Matter}},}\ }\href {\doibase 10.1103/PhysRevD.86.075005} {\bibfield  {journal} {\bibinfo  {journal} {Phys. Rev. D}\ }\textbf {\bibinfo {volume} {86}},\ \bibinfo {pages} {075005} (\bibinfo {year} {2012})},\ \Eprint {http://arxiv.org/abs/1207.6388} {arXiv:1207.6388 [hep-ph]} \BibitemShut {NoStop}%
\bibitem [{\citenamefont {Dorsner}\ \emph {et~al.}(2014)\citenamefont {Dorsner}, \citenamefont {Fajfer},\ and\ \citenamefont {Mustac}}]{Dorsner:2014wva}%
  \BibitemOpen
  \bibfield  {author} {\bibinfo {author} {\bibfnamefont {Ilja}\ \bibnamefont {Dorsner}}, \bibinfo {author} {\bibfnamefont {Svjetlana}\ \bibnamefont {Fajfer}}, \ and\ \bibinfo {author} {\bibfnamefont {Ivana}\ \bibnamefont {Mustac}},\ }\bibfield  {title} {\enquote {\bibinfo {title} {{Light vector-like fermions in a minimal SU(5) setup}},}\ }\href {\doibase 10.1103/PhysRevD.89.115004} {\bibfield  {journal} {\bibinfo  {journal} {Phys. Rev. D}\ }\textbf {\bibinfo {volume} {89}},\ \bibinfo {pages} {115004} (\bibinfo {year} {2014})},\ \Eprint {http://arxiv.org/abs/1401.6870} {arXiv:1401.6870 [hep-ph]} \BibitemShut {NoStop}%
\bibitem [{\citenamefont {Fileviez~P\'erez}\ \emph {et~al.}(2018)\citenamefont {Fileviez~P\'erez}, \citenamefont {Gross},\ and\ \citenamefont {Murgui}}]{FileviezPerez:2018dyf}%
  \BibitemOpen
  \bibfield  {author} {\bibinfo {author} {\bibfnamefont {Pavel}\ \bibnamefont {Fileviez~P\'erez}}, \bibinfo {author} {\bibfnamefont {Axel}\ \bibnamefont {Gross}}, \ and\ \bibinfo {author} {\bibfnamefont {Clara}\ \bibnamefont {Murgui}},\ }\bibfield  {title} {\enquote {\bibinfo {title} {{Seesaw scale, unification, and proton decay}},}\ }\href {\doibase 10.1103/PhysRevD.98.035032} {\bibfield  {journal} {\bibinfo  {journal} {Phys. Rev. D}\ }\textbf {\bibinfo {volume} {98}},\ \bibinfo {pages} {035032} (\bibinfo {year} {2018})},\ \Eprint {http://arxiv.org/abs/1804.07831} {arXiv:1804.07831 [hep-ph]} \BibitemShut {NoStop}%
\bibitem [{\citenamefont {Antusch}\ \emph {et~al.}(2023{\natexlab{b}})\citenamefont {Antusch}, \citenamefont {Hinze}, \citenamefont {Saad},\ and\ \citenamefont {Steiner}}]{Antusch:2023mxx}%
  \BibitemOpen
  \bibfield  {author} {\bibinfo {author} {\bibfnamefont {Stefan}\ \bibnamefont {Antusch}}, \bibinfo {author} {\bibfnamefont {Kevin}\ \bibnamefont {Hinze}}, \bibinfo {author} {\bibfnamefont {Shaikh}\ \bibnamefont {Saad}}, \ and\ \bibinfo {author} {\bibfnamefont {Jonathan}\ \bibnamefont {Steiner}},\ }\bibfield  {title} {\enquote {\bibinfo {title} {{A Generalised Missing Partner Mechanism for SU(5) GUT Inflation}},}\ }\href@noop {} {\  (\bibinfo {year} {2023}{\natexlab{b}})},\ \Eprint {http://arxiv.org/abs/2308.11705} {arXiv:2308.11705 [hep-ph]} \BibitemShut {NoStop}%
\bibitem [{\citenamefont {Antusch}\ \emph {et~al.}(2023{\natexlab{c}})\citenamefont {Antusch}, \citenamefont {Hinze},\ and\ \citenamefont {Saad}}]{Antusch:2023kli}%
  \BibitemOpen
  \bibfield  {author} {\bibinfo {author} {\bibfnamefont {Stefan}\ \bibnamefont {Antusch}}, \bibinfo {author} {\bibfnamefont {Kevin}\ \bibnamefont {Hinze}}, \ and\ \bibinfo {author} {\bibfnamefont {Shaikh}\ \bibnamefont {Saad}},\ }\bibfield  {title} {\enquote {\bibinfo {title} {{Quark-lepton Yukawa ratios and nucleon decay in SU(5) GUTs with type-III seesaw}},}\ }\href {\doibase 10.1016/j.nuclphysb.2023.116195} {\bibfield  {journal} {\bibinfo  {journal} {Nucl. Phys. B}\ }\textbf {\bibinfo {volume} {991}},\ \bibinfo {pages} {116195} (\bibinfo {year} {2023}{\natexlab{c}})},\ \Eprint {http://arxiv.org/abs/2301.03601} {arXiv:2301.03601 [hep-ph]} \BibitemShut {NoStop}%
\bibitem [{\citenamefont {Antusch}\ \emph {et~al.}(2023{\natexlab{d}})\citenamefont {Antusch}, \citenamefont {Hinze},\ and\ \citenamefont {Saad}}]{Antusch:2023mqe}%
  \BibitemOpen
  \bibfield  {author} {\bibinfo {author} {\bibfnamefont {Stefan}\ \bibnamefont {Antusch}}, \bibinfo {author} {\bibfnamefont {Kevin}\ \bibnamefont {Hinze}}, \ and\ \bibinfo {author} {\bibfnamefont {Shaikh}\ \bibnamefont {Saad}},\ }\bibfield  {title} {\enquote {\bibinfo {title} {{Minimal $SU(5)$ GUTs with vectorlike fermions}},}\ }\href@noop {} {\  (\bibinfo {year} {2023}{\natexlab{d}})},\ \Eprint {http://arxiv.org/abs/2308.08585} {arXiv:2308.08585 [hep-ph]} \BibitemShut {NoStop}%
\bibitem [{\citenamefont {Patel}\ and\ \citenamefont {Shukla}(2024)}]{Patel:2023gwt}%
  \BibitemOpen
  \bibfield  {author} {\bibinfo {author} {\bibfnamefont {Ketan~M.}\ \bibnamefont {Patel}}\ and\ \bibinfo {author} {\bibfnamefont {Saurabh~K.}\ \bibnamefont {Shukla}},\ }\bibfield  {title} {\enquote {\bibinfo {title} {{Quantum corrections and the minimal Yukawa sector of SU(5)}},}\ }\href {\doibase 10.1103/PhysRevD.109.015007} {\bibfield  {journal} {\bibinfo  {journal} {Phys. Rev. D}\ }\textbf {\bibinfo {volume} {109}},\ \bibinfo {pages} {015007} (\bibinfo {year} {2024})},\ \Eprint {http://arxiv.org/abs/2310.16563} {arXiv:2310.16563 [hep-ph]} \BibitemShut {NoStop}%
\bibitem [{\citenamefont {del Aguila}\ and\ \citenamefont {Ibanez}(1981)}]{delAguila:1980qag}%
  \BibitemOpen
  \bibfield  {author} {\bibinfo {author} {\bibfnamefont {F.}~\bibnamefont {del Aguila}}\ and\ \bibinfo {author} {\bibfnamefont {Luis~E.}\ \bibnamefont {Ibanez}},\ }\bibfield  {title} {\enquote {\bibinfo {title} {{Higgs Bosons in SO(10) and Partial Unification}},}\ }\href {\doibase 10.1016/0550-3213(81)90266-2} {\bibfield  {journal} {\bibinfo  {journal} {Nucl. Phys. B}\ }\textbf {\bibinfo {volume} {177}},\ \bibinfo {pages} {60--86} (\bibinfo {year} {1981})}\BibitemShut {NoStop}%
\bibitem [{\citenamefont {Mohapatra}\ and\ \citenamefont {Senjanovic}(1983)}]{Mohapatra:1982aq}%
  \BibitemOpen
  \bibfield  {author} {\bibinfo {author} {\bibfnamefont {Rabindra~N.}\ \bibnamefont {Mohapatra}}\ and\ \bibinfo {author} {\bibfnamefont {Goran}\ \bibnamefont {Senjanovic}},\ }\bibfield  {title} {\enquote {\bibinfo {title} {{Higgs Boson Effects in Grand Unified Theories}},}\ }\href {\doibase 10.1103/PhysRevD.27.1601} {\bibfield  {journal} {\bibinfo  {journal} {Phys. Rev. D}\ }\textbf {\bibinfo {volume} {27}},\ \bibinfo {pages} {1601} (\bibinfo {year} {1983})}\BibitemShut {NoStop}%
\bibitem [{\citenamefont {Dimopoulos}\ and\ \citenamefont {Georgi}(1984)}]{Dimopoulos:1984ha}%
  \BibitemOpen
  \bibfield  {author} {\bibinfo {author} {\bibfnamefont {S.}~\bibnamefont {Dimopoulos}}\ and\ \bibinfo {author} {\bibfnamefont {H.~M.}\ \bibnamefont {Georgi}},\ }\bibfield  {title} {\enquote {\bibinfo {title} {{Extended Survival Hypothesis and Fermion Masses}},}\ }\href {\doibase 10.1016/0370-2693(84)91049-9} {\bibfield  {journal} {\bibinfo  {journal} {Phys. Lett. B}\ }\textbf {\bibinfo {volume} {140}},\ \bibinfo {pages} {67--70} (\bibinfo {year} {1984})}\BibitemShut {NoStop}%
\bibitem [{\citenamefont {Babu}\ and\ \citenamefont {Ma}(1984)}]{Babu:1984vx}%
  \BibitemOpen
  \bibfield  {author} {\bibinfo {author} {\bibfnamefont {K.~S.}\ \bibnamefont {Babu}}\ and\ \bibinfo {author} {\bibfnamefont {Ernest}\ \bibnamefont {Ma}},\ }\bibfield  {title} {\enquote {\bibinfo {title} {{Suppression of Proton Decay in SU(5) Grand Unification}},}\ }\href {\doibase 10.1016/0370-2693(84)91283-8} {\bibfield  {journal} {\bibinfo  {journal} {Phys. Lett. B}\ }\textbf {\bibinfo {volume} {144}},\ \bibinfo {pages} {381--385} (\bibinfo {year} {1984})}\BibitemShut {NoStop}%
\bibitem [{\citenamefont {Fileviez~Perez}\ and\ \citenamefont {Murgui}(2016)}]{FileviezPerez:2016sal}%
  \BibitemOpen
  \bibfield  {author} {\bibinfo {author} {\bibfnamefont {Pavel}\ \bibnamefont {Fileviez~Perez}}\ and\ \bibinfo {author} {\bibfnamefont {Clara}\ \bibnamefont {Murgui}},\ }\bibfield  {title} {\enquote {\bibinfo {title} {{Renormalizable SU(5) Unification}},}\ }\href {\doibase 10.1103/PhysRevD.94.075014} {\bibfield  {journal} {\bibinfo  {journal} {Phys. Rev. D}\ }\textbf {\bibinfo {volume} {94}},\ \bibinfo {pages} {075014} (\bibinfo {year} {2016})},\ \Eprint {http://arxiv.org/abs/1604.03377} {arXiv:1604.03377 [hep-ph]} \BibitemShut {NoStop}%
\bibitem [{\citenamefont {Haba}\ \emph {et~al.}(2024)\citenamefont {Haba}, \citenamefont {Nagano}, \citenamefont {Shimizu},\ and\ \citenamefont {Yamada}}]{Haba:2024lox}%
  \BibitemOpen
  \bibfield  {author} {\bibinfo {author} {\bibfnamefont {Naoyuki}\ \bibnamefont {Haba}}, \bibinfo {author} {\bibfnamefont {Keisuke}\ \bibnamefont {Nagano}}, \bibinfo {author} {\bibfnamefont {Yasuhiro}\ \bibnamefont {Shimizu}}, \ and\ \bibinfo {author} {\bibfnamefont {Toshifumi}\ \bibnamefont {Yamada}},\ }\bibfield  {title} {\enquote {\bibinfo {title} {{Gauge Coupling Unification and Proton Decay via 45 Higgs Boson in SU(5) GUT}},}\ }\href {\doibase 10.1093/ptep/ptae066} {\bibfield  {journal} {\bibinfo  {journal} {PTEP}\ }\textbf {\bibinfo {volume} {2024}},\ \bibinfo {pages} {053B05} (\bibinfo {year} {2024})},\ \Eprint {http://arxiv.org/abs/2402.15124} {arXiv:2402.15124 [hep-ph]} \BibitemShut {NoStop}%
\bibitem [{\citenamefont {Fang}\ and\ \citenamefont {Zhou}(2024)}]{Fang:2024mfn}%
  \BibitemOpen
  \bibfield  {author} {\bibinfo {author} {\bibfnamefont {Gao-Xiang}\ \bibnamefont {Fang}}\ and\ \bibinfo {author} {\bibfnamefont {Ye-Ling}\ \bibnamefont {Zhou}},\ }\bibfield  {title} {\enquote {\bibinfo {title} {{Exploring flavour space of an economical SU(5) GUT in future proton decay measurements}},}\ }\href@noop {} {\  (\bibinfo {year} {2024})},\ \Eprint {http://arxiv.org/abs/2406.06861} {arXiv:2406.06861 [hep-ph]} \BibitemShut {NoStop}%
\bibitem [{\citenamefont {Be\v{c}irevi\'c}\ \emph {et~al.}(2018)\citenamefont {Be\v{c}irevi\'c}, \citenamefont {Dor\v{s}ner}, \citenamefont {Fajfer}, \citenamefont {Ko\v{s}nik}, \citenamefont {Faroughy},\ and\ \citenamefont {Sumensari}}]{Becirevic:2018afm}%
  \BibitemOpen
  \bibfield  {author} {\bibinfo {author} {\bibfnamefont {Damir}\ \bibnamefont {Be\v{c}irevi\'c}}, \bibinfo {author} {\bibfnamefont {Ilja}\ \bibnamefont {Dor\v{s}ner}}, \bibinfo {author} {\bibfnamefont {Svjetlana}\ \bibnamefont {Fajfer}}, \bibinfo {author} {\bibfnamefont {Nejc}\ \bibnamefont {Ko\v{s}nik}}, \bibinfo {author} {\bibfnamefont {Darius~A.}\ \bibnamefont {Faroughy}}, \ and\ \bibinfo {author} {\bibfnamefont {Olcyr}\ \bibnamefont {Sumensari}},\ }\bibfield  {title} {\enquote {\bibinfo {title} {{Scalar leptoquarks from grand unified theories to accommodate the $B$-physics anomalies}},}\ }\href {\doibase 10.1103/PhysRevD.98.055003} {\bibfield  {journal} {\bibinfo  {journal} {Phys. Rev. D}\ }\textbf {\bibinfo {volume} {98}},\ \bibinfo {pages} {055003} (\bibinfo {year} {2018})},\ \Eprint {http://arxiv.org/abs/1806.05689} {arXiv:1806.05689 [hep-ph]} \BibitemShut {NoStop}%
\bibitem [{\citenamefont {Dorsner}\ and\ \citenamefont {Mocioiu}(2008)}]{Dorsner:2007fy}%
  \BibitemOpen
  \bibfield  {author} {\bibinfo {author} {\bibfnamefont {Ilja}\ \bibnamefont {Dorsner}}\ and\ \bibinfo {author} {\bibfnamefont {Irina}\ \bibnamefont {Mocioiu}},\ }\bibfield  {title} {\enquote {\bibinfo {title} {{Predictions from type II see-saw mechanism in SU(5)}},}\ }\href {\doibase 10.1016/j.nuclphysb.2007.12.004} {\bibfield  {journal} {\bibinfo  {journal} {Nucl. Phys. B}\ }\textbf {\bibinfo {volume} {796}},\ \bibinfo {pages} {123--136} (\bibinfo {year} {2008})},\ \Eprint {http://arxiv.org/abs/0708.3332} {arXiv:0708.3332 [hep-ph]} \BibitemShut {NoStop}%
\bibitem [{\citenamefont {Dor\v{s}ner}\ \emph {et~al.}(2017)\citenamefont {Dor\v{s}ner}, \citenamefont {Fajfer},\ and\ \citenamefont {Ko\v{s}nik}}]{Dorsner:2017wwn}%
  \BibitemOpen
  \bibfield  {author} {\bibinfo {author} {\bibfnamefont {Ilja}\ \bibnamefont {Dor\v{s}ner}}, \bibinfo {author} {\bibfnamefont {Svjetlana}\ \bibnamefont {Fajfer}}, \ and\ \bibinfo {author} {\bibfnamefont {Nejc}\ \bibnamefont {Ko\v{s}nik}},\ }\bibfield  {title} {\enquote {\bibinfo {title} {{Leptoquark mechanism of neutrino masses within the grand unification framework}},}\ }\href {\doibase 10.1140/epjc/s10052-017-4987-2} {\bibfield  {journal} {\bibinfo  {journal} {Eur. Phys. J. C}\ }\textbf {\bibinfo {volume} {77}},\ \bibinfo {pages} {417} (\bibinfo {year} {2017})},\ \Eprint {http://arxiv.org/abs/1701.08322} {arXiv:1701.08322 [hep-ph]} \BibitemShut {NoStop}%
\bibitem [{\citenamefont {Schechter}\ and\ \citenamefont {Valle}(1980)}]{Schechter:1980gr}%
  \BibitemOpen
  \bibfield  {author} {\bibinfo {author} {\bibfnamefont {J.}~\bibnamefont {Schechter}}\ and\ \bibinfo {author} {\bibfnamefont {J.~W.~F.}\ \bibnamefont {Valle}},\ }\bibfield  {title} {\enquote {\bibinfo {title} {{Neutrino Masses in SU(2) x U(1) Theories}},}\ }\href {\doibase 10.1103/PhysRevD.22.2227} {\bibfield  {journal} {\bibinfo  {journal} {Phys. Rev. D}\ }\textbf {\bibinfo {volume} {22}},\ \bibinfo {pages} {2227} (\bibinfo {year} {1980})}\BibitemShut {NoStop}%
\bibitem [{\citenamefont {Mohapatra}\ and\ \citenamefont {Senjanovic}(1981)}]{Mohapatra:1980yp}%
  \BibitemOpen
  \bibfield  {author} {\bibinfo {author} {\bibfnamefont {Rabindra~N.}\ \bibnamefont {Mohapatra}}\ and\ \bibinfo {author} {\bibfnamefont {Goran}\ \bibnamefont {Senjanovic}},\ }\bibfield  {title} {\enquote {\bibinfo {title} {{Neutrino Masses and Mixings in Gauge Models with Spontaneous Parity Violation}},}\ }\href {\doibase 10.1103/PhysRevD.23.165} {\bibfield  {journal} {\bibinfo  {journal} {Phys. Rev. D}\ }\textbf {\bibinfo {volume} {23}},\ \bibinfo {pages} {165} (\bibinfo {year} {1981})}\BibitemShut {NoStop}%
\bibitem [{\citenamefont {Slansky}(1981)}]{Slansky:1981yr}%
  \BibitemOpen
  \bibfield  {author} {\bibinfo {author} {\bibfnamefont {R.}~\bibnamefont {Slansky}},\ }\bibfield  {title} {\enquote {\bibinfo {title} {{Group Theory for Unified Model Building}},}\ }\href {\doibase 10.1016/0370-1573(81)90092-2} {\bibfield  {journal} {\bibinfo  {journal} {Phys. Rept.}\ }\textbf {\bibinfo {volume} {79}},\ \bibinfo {pages} {1--128} (\bibinfo {year} {1981})}\BibitemShut {NoStop}%
\bibitem [{\citenamefont {Weinberg}(1980)}]{Weinberg:1980wa}%
  \BibitemOpen
  \bibfield  {author} {\bibinfo {author} {\bibfnamefont {Steven}\ \bibnamefont {Weinberg}},\ }\bibfield  {title} {\enquote {\bibinfo {title} {{Effective Gauge Theories}},}\ }\href {\doibase 10.1016/0370-2693(80)90660-7} {\bibfield  {journal} {\bibinfo  {journal} {Phys. Lett. B}\ }\textbf {\bibinfo {volume} {91}},\ \bibinfo {pages} {51--55} (\bibinfo {year} {1980})}\BibitemShut {NoStop}%
\bibitem [{\citenamefont {Hall}(1981)}]{Hall:1980kf}%
  \BibitemOpen
  \bibfield  {author} {\bibinfo {author} {\bibfnamefont {Lawrence~J.}\ \bibnamefont {Hall}},\ }\bibfield  {title} {\enquote {\bibinfo {title} {{Grand Unification of Effective Gauge Theories}},}\ }\href {\doibase 10.1016/0550-3213(81)90498-3} {\bibfield  {journal} {\bibinfo  {journal} {Nucl. Phys. B}\ }\textbf {\bibinfo {volume} {178}},\ \bibinfo {pages} {75--124} (\bibinfo {year} {1981})}\BibitemShut {NoStop}%
\bibitem [{\citenamefont {Oliensis}\ and\ \citenamefont {Fischler}(1983)}]{PhysRevD.28.194}%
  \BibitemOpen
  \bibfield  {author} {\bibinfo {author} {\bibfnamefont {J.}~\bibnamefont {Oliensis}}\ and\ \bibinfo {author} {\bibfnamefont {M.}~\bibnamefont {Fischler}},\ }\bibfield  {title} {\enquote {\bibinfo {title} {Two-loop calculations of $\frac{{M}_{b}}{{M}_{\ensuremath{\tau}}}$ and heavy-fermion masses in the su(5) model},}\ }\href {\doibase 10.1103/PhysRevD.28.194} {\bibfield  {journal} {\bibinfo  {journal} {Phys. Rev. D}\ }\textbf {\bibinfo {volume} {28}},\ \bibinfo {pages} {194--199} (\bibinfo {year} {1983})}\BibitemShut {NoStop}%
\bibitem [{\citenamefont {Kane}\ \emph {et~al.}(1994)\citenamefont {Kane}, \citenamefont {Kolda}, \citenamefont {Roszkowski},\ and\ \citenamefont {Wells}}]{Kane:1993td}%
  \BibitemOpen
  \bibfield  {author} {\bibinfo {author} {\bibfnamefont {Gordon~L.}\ \bibnamefont {Kane}}, \bibinfo {author} {\bibfnamefont {Christopher~F.}\ \bibnamefont {Kolda}}, \bibinfo {author} {\bibfnamefont {Leszek}\ \bibnamefont {Roszkowski}}, \ and\ \bibinfo {author} {\bibfnamefont {James~D.}\ \bibnamefont {Wells}},\ }\bibfield  {title} {\enquote {\bibinfo {title} {{Study of constrained minimal supersymmetry}},}\ }\href {\doibase 10.1103/PhysRevD.49.6173} {\bibfield  {journal} {\bibinfo  {journal} {Phys. Rev. D}\ }\textbf {\bibinfo {volume} {49}},\ \bibinfo {pages} {6173--6210} (\bibinfo {year} {1994})},\ \Eprint {http://arxiv.org/abs/hep-ph/9312272} {arXiv:hep-ph/9312272} \BibitemShut {NoStop}%
\bibitem [{\citenamefont {Wright}(1994)}]{Wright:1994qb}%
  \BibitemOpen
  \bibfield  {author} {\bibinfo {author} {\bibfnamefont {Brian~D.}\ \bibnamefont {Wright}},\ }\bibfield  {title} {\enquote {\bibinfo {title} {{Yukawa coupling thresholds: Application to the MSSM and the minimal supersymmetric SU(5) GUT}},}\ }\href@noop {} {\  (\bibinfo {year} {1994})},\ \Eprint {http://arxiv.org/abs/hep-ph/9404217} {arXiv:hep-ph/9404217} \BibitemShut {NoStop}%
\bibitem [{\citenamefont {Patel}\ and\ \citenamefont {Shukla}(2022)}]{Patel:2022wya}%
  \BibitemOpen
  \bibfield  {author} {\bibinfo {author} {\bibfnamefont {Ketan~M.}\ \bibnamefont {Patel}}\ and\ \bibinfo {author} {\bibfnamefont {Saurabh~K.}\ \bibnamefont {Shukla}},\ }\bibfield  {title} {\enquote {\bibinfo {title} {{Anatomy of scalar mediated proton decays in SO(10) models}},}\ }\href {\doibase 10.1007/JHEP08(2022)042} {\bibfield  {journal} {\bibinfo  {journal} {JHEP}\ }\textbf {\bibinfo {volume} {08}},\ \bibinfo {pages} {042} (\bibinfo {year} {2022})},\ \Eprint {http://arxiv.org/abs/2203.07748} {arXiv:2203.07748 [hep-ph]} \BibitemShut {NoStop}%
\bibitem [{\citenamefont {Buras}\ \emph {et~al.}(1978)\citenamefont {Buras}, \citenamefont {Ellis}, \citenamefont {Gaillard},\ and\ \citenamefont {Nanopoulos}}]{Buras:1977yy}%
  \BibitemOpen
  \bibfield  {author} {\bibinfo {author} {\bibfnamefont {A.~J.}\ \bibnamefont {Buras}}, \bibinfo {author} {\bibfnamefont {John~R.}\ \bibnamefont {Ellis}}, \bibinfo {author} {\bibfnamefont {M.~K.}\ \bibnamefont {Gaillard}}, \ and\ \bibinfo {author} {\bibfnamefont {Dimitri~V.}\ \bibnamefont {Nanopoulos}},\ }\bibfield  {title} {\enquote {\bibinfo {title} {{Aspects of the Grand Unification of Strong, Weak and Electromagnetic Interactions}},}\ }\href {\doibase 10.1016/0550-3213(78)90214-6} {\bibfield  {journal} {\bibinfo  {journal} {Nucl. Phys. B}\ }\textbf {\bibinfo {volume} {135}},\ \bibinfo {pages} {66--92} (\bibinfo {year} {1978})}\BibitemShut {NoStop}%
\bibitem [{\citenamefont {Dor\v{s}ner}\ \emph {et~al.}(2016)\citenamefont {Dor\v{s}ner}, \citenamefont {Fajfer}, \citenamefont {Greljo}, \citenamefont {Kamenik},\ and\ \citenamefont {Ko\v{s}nik}}]{Dorsner:2016wpm}%
  \BibitemOpen
  \bibfield  {author} {\bibinfo {author} {\bibfnamefont {I.}~\bibnamefont {Dor\v{s}ner}}, \bibinfo {author} {\bibfnamefont {S.}~\bibnamefont {Fajfer}}, \bibinfo {author} {\bibfnamefont {A.}~\bibnamefont {Greljo}}, \bibinfo {author} {\bibfnamefont {J.~F.}\ \bibnamefont {Kamenik}}, \ and\ \bibinfo {author} {\bibfnamefont {N.}~\bibnamefont {Ko\v{s}nik}},\ }\bibfield  {title} {\enquote {\bibinfo {title} {{Physics of leptoquarks in precision experiments and at particle colliders}},}\ }\href {\doibase 10.1016/j.physrep.2016.06.001} {\bibfield  {journal} {\bibinfo  {journal} {Phys. Rept.}\ }\textbf {\bibinfo {volume} {641}},\ \bibinfo {pages} {1--68} (\bibinfo {year} {2016})},\ \Eprint {http://arxiv.org/abs/1603.04993} {arXiv:1603.04993 [hep-ph]} \BibitemShut {NoStop}%
\bibitem [{\citenamefont {Allwicher}\ \emph {et~al.}(2021)\citenamefont {Allwicher}, \citenamefont {Arnan}, \citenamefont {Barducci},\ and\ \citenamefont {Nardecchia}}]{Allwicher:2021rtd}%
  \BibitemOpen
  \bibfield  {author} {\bibinfo {author} {\bibfnamefont {Lukas}\ \bibnamefont {Allwicher}}, \bibinfo {author} {\bibfnamefont {Pere}\ \bibnamefont {Arnan}}, \bibinfo {author} {\bibfnamefont {Daniele}\ \bibnamefont {Barducci}}, \ and\ \bibinfo {author} {\bibfnamefont {Marco}\ \bibnamefont {Nardecchia}},\ }\bibfield  {title} {\enquote {\bibinfo {title} {{Perturbative unitarity constraints on generic Yukawa interactions}},}\ }\href {\doibase 10.1007/JHEP10(2021)129} {\bibfield  {journal} {\bibinfo  {journal} {JHEP}\ }\textbf {\bibinfo {volume} {10}},\ \bibinfo {pages} {129} (\bibinfo {year} {2021})},\ \Eprint {http://arxiv.org/abs/2108.00013} {arXiv:2108.00013 [hep-ph]} \BibitemShut {NoStop}%
\bibitem [{\citenamefont {Joshipura}\ and\ \citenamefont {Patel}(2011)}]{Joshipura:2011nn}%
  \BibitemOpen
  \bibfield  {author} {\bibinfo {author} {\bibfnamefont {Anjan~S.}\ \bibnamefont {Joshipura}}\ and\ \bibinfo {author} {\bibfnamefont {Ketan~M.}\ \bibnamefont {Patel}},\ }\bibfield  {title} {\enquote {\bibinfo {title} {{Fermion Masses in SO(10) Models}},}\ }\href {\doibase 10.1103/PhysRevD.83.095002} {\bibfield  {journal} {\bibinfo  {journal} {Phys. Rev. D}\ }\textbf {\bibinfo {volume} {83}},\ \bibinfo {pages} {095002} (\bibinfo {year} {2011})},\ \Eprint {http://arxiv.org/abs/1102.5148} {arXiv:1102.5148 [hep-ph]} \BibitemShut {NoStop}%
\bibitem [{\citenamefont {Mummidi}\ and\ \citenamefont {Patel}(2021)}]{Mummidi:2021anm}%
  \BibitemOpen
  \bibfield  {author} {\bibinfo {author} {\bibfnamefont {V.~Suryanarayana}\ \bibnamefont {Mummidi}}\ and\ \bibinfo {author} {\bibfnamefont {Ketan~M.}\ \bibnamefont {Patel}},\ }\bibfield  {title} {\enquote {\bibinfo {title} {{Leptogenesis and fermion mass fit in a renormalizable SO(10) model}},}\ }\href {\doibase 10.1007/JHEP12(2021)042} {\bibfield  {journal} {\bibinfo  {journal} {JHEP}\ }\textbf {\bibinfo {volume} {12}},\ \bibinfo {pages} {042} (\bibinfo {year} {2021})},\ \Eprint {http://arxiv.org/abs/2109.04050} {arXiv:2109.04050 [hep-ph]} \BibitemShut {NoStop}%
\bibitem [{\citenamefont {Sartore}\ and\ \citenamefont {Schienbein}(2021)}]{Sartore:2020gou}%
  \BibitemOpen
  \bibfield  {author} {\bibinfo {author} {\bibfnamefont {Lohan}\ \bibnamefont {Sartore}}\ and\ \bibinfo {author} {\bibfnamefont {Ingo}\ \bibnamefont {Schienbein}},\ }\bibfield  {title} {\enquote {\bibinfo {title} {{PyR@TE 3}},}\ }\href {\doibase 10.1016/j.cpc.2020.107819} {\bibfield  {journal} {\bibinfo  {journal} {Comput. Phys. Commun.}\ }\textbf {\bibinfo {volume} {261}},\ \bibinfo {pages} {107819} (\bibinfo {year} {2021})},\ \Eprint {http://arxiv.org/abs/2007.12700} {arXiv:2007.12700 [hep-ph]} \BibitemShut {NoStop}%
\bibitem [{\citenamefont {Navas}\ \emph {et~al.}(2024)\citenamefont {Navas} \emph {et~al.}}]{ParticleDataGroup:2024cfk}%
  \BibitemOpen
  \bibfield  {author} {\bibinfo {author} {\bibfnamefont {S.}~\bibnamefont {Navas}} \emph {et~al.} (\bibinfo {collaboration} {Particle Data Group}),\ }\bibfield  {title} {\enquote {\bibinfo {title} {{Review of particle physics}},}\ }\href {\doibase 10.1103/PhysRevD.110.030001} {\bibfield  {journal} {\bibinfo  {journal} {Phys. Rev. D}\ }\textbf {\bibinfo {volume} {110}},\ \bibinfo {pages} {030001} (\bibinfo {year} {2024})}\BibitemShut {NoStop}%
\bibitem [{\citenamefont {Babu}\ \emph {et~al.}(2017)\citenamefont {Babu}, \citenamefont {Bajc},\ and\ \citenamefont {Saad}}]{Babu:2016bmy}%
  \BibitemOpen
  \bibfield  {author} {\bibinfo {author} {\bibfnamefont {K.~S.}\ \bibnamefont {Babu}}, \bibinfo {author} {\bibfnamefont {Borut}\ \bibnamefont {Bajc}}, \ and\ \bibinfo {author} {\bibfnamefont {Shaikh}\ \bibnamefont {Saad}},\ }\bibfield  {title} {\enquote {\bibinfo {title} {{Yukawa Sector of Minimal SO(10) Unification}},}\ }\href {\doibase 10.1007/JHEP02(2017)136} {\bibfield  {journal} {\bibinfo  {journal} {JHEP}\ }\textbf {\bibinfo {volume} {02}},\ \bibinfo {pages} {136} (\bibinfo {year} {2017})},\ \Eprint {http://arxiv.org/abs/1612.04329} {arXiv:1612.04329 [hep-ph]} \BibitemShut {NoStop}%
\bibitem [{\citenamefont {Dueck}\ and\ \citenamefont {Rodejohann}(2013)}]{Dueck:2013gca}%
  \BibitemOpen
  \bibfield  {author} {\bibinfo {author} {\bibfnamefont {Alexander}\ \bibnamefont {Dueck}}\ and\ \bibinfo {author} {\bibfnamefont {Werner}\ \bibnamefont {Rodejohann}},\ }\bibfield  {title} {\enquote {\bibinfo {title} {{Fits to SO(10) Grand Unified Models}},}\ }\href {\doibase 10.1007/JHEP09(2013)024} {\bibfield  {journal} {\bibinfo  {journal} {JHEP}\ }\textbf {\bibinfo {volume} {09}},\ \bibinfo {pages} {024} (\bibinfo {year} {2013})},\ \Eprint {http://arxiv.org/abs/1306.4468} {arXiv:1306.4468 [hep-ph]} \BibitemShut {NoStop}%
\bibitem [{\citenamefont {Dreiner}\ \emph {et~al.}(2010)\citenamefont {Dreiner}, \citenamefont {Haber},\ and\ \citenamefont {Martin}}]{Dreiner:2008tw}%
  \BibitemOpen
  \bibfield  {author} {\bibinfo {author} {\bibfnamefont {Herbi~K.}\ \bibnamefont {Dreiner}}, \bibinfo {author} {\bibfnamefont {Howard~E.}\ \bibnamefont {Haber}}, \ and\ \bibinfo {author} {\bibfnamefont {Stephen~P.}\ \bibnamefont {Martin}},\ }\bibfield  {title} {\enquote {\bibinfo {title} {{Two-component spinor techniques and Feynman rules for quantum field theory and supersymmetry}},}\ }\href {\doibase 10.1016/j.physrep.2010.05.002} {\bibfield  {journal} {\bibinfo  {journal} {Phys. Rept.}\ }\textbf {\bibinfo {volume} {494}},\ \bibinfo {pages} {1--196} (\bibinfo {year} {2010})},\ \Eprint {http://arxiv.org/abs/0812.1594} {arXiv:0812.1594 [hep-ph]} \BibitemShut {NoStop}%
\bibitem [{\citenamefont {Abe}\ \emph {et~al.}(2014{\natexlab{a}})\citenamefont {Abe} \emph {et~al.}}]{Super-Kamiokande:2013rwg}%
  \BibitemOpen
  \bibfield  {author} {\bibinfo {author} {\bibfnamefont {K.}~\bibnamefont {Abe}} \emph {et~al.} (\bibinfo {collaboration} {Super-Kamiokande}),\ }\bibfield  {title} {\enquote {\bibinfo {title} {{Search for Nucleon Decay via $n \to \bar{\nu} \pi^{0}$ and $p \to \bar{\nu} \pi^{+}$ in Super-Kamiokande}},}\ }\href {\doibase 10.1103/PhysRevLett.113.121802} {\bibfield  {journal} {\bibinfo  {journal} {Phys. Rev. Lett.}\ }\textbf {\bibinfo {volume} {113}},\ \bibinfo {pages} {121802} (\bibinfo {year} {2014}{\natexlab{a}})},\ \Eprint {http://arxiv.org/abs/1305.4391} {arXiv:1305.4391 [hep-ex]} \BibitemShut {NoStop}%
\bibitem [{\citenamefont {Abe}\ \emph {et~al.}(2014{\natexlab{b}})\citenamefont {Abe} \emph {et~al.}}]{Super-Kamiokande:2014otb}%
  \BibitemOpen
  \bibfield  {author} {\bibinfo {author} {\bibfnamefont {K.}~\bibnamefont {Abe}} \emph {et~al.} (\bibinfo {collaboration} {Super-Kamiokande}),\ }\bibfield  {title} {\enquote {\bibinfo {title} {{Search for proton decay via $p\to\nu K^+$ using 260 kiloton\textperiodcentered{}year data of Super-Kamiokande}},}\ }\href {\doibase 10.1103/PhysRevD.90.072005} {\bibfield  {journal} {\bibinfo  {journal} {Phys. Rev. D}\ }\textbf {\bibinfo {volume} {90}},\ \bibinfo {pages} {072005} (\bibinfo {year} {2014}{\natexlab{b}})},\ \Eprint {http://arxiv.org/abs/1408.1195} {arXiv:1408.1195 [hep-ex]} \BibitemShut {NoStop}%
\bibitem [{\citenamefont {Esteban}\ \emph {et~al.}(2024)\citenamefont {Esteban}, \citenamefont {Gonzalez-Garcia}, \citenamefont {Maltoni}, \citenamefont {Martinez-Soler}, \citenamefont {Pinheiro},\ and\ \citenamefont {Schwetz}}]{Esteban:2024eli}%
  \BibitemOpen
  \bibfield  {author} {\bibinfo {author} {\bibfnamefont {Ivan}\ \bibnamefont {Esteban}}, \bibinfo {author} {\bibfnamefont {M.~C.}\ \bibnamefont {Gonzalez-Garcia}}, \bibinfo {author} {\bibfnamefont {Michele}\ \bibnamefont {Maltoni}}, \bibinfo {author} {\bibfnamefont {Ivan}\ \bibnamefont {Martinez-Soler}}, \bibinfo {author} {\bibfnamefont {Jo\~ao~Paulo}\ \bibnamefont {Pinheiro}}, \ and\ \bibinfo {author} {\bibfnamefont {Thomas}\ \bibnamefont {Schwetz}},\ }\bibfield  {title} {\enquote {\bibinfo {title} {{NuFit-6.0: Updated global analysis of three-flavor neutrino oscillations}},}\ }\href@noop {} {\  (\bibinfo {year} {2024})},\ \Eprint {http://arxiv.org/abs/2410.05380} {arXiv:2410.05380 [hep-ph]} \BibitemShut {NoStop}%
\bibitem [{\citenamefont {Chankowski}\ and\ \citenamefont {Pluciennik}(1993)}]{Chankowski:1993tx}%
  \BibitemOpen
  \bibfield  {author} {\bibinfo {author} {\bibfnamefont {Piotr~H.}\ \bibnamefont {Chankowski}}\ and\ \bibinfo {author} {\bibfnamefont {Zbigniew}\ \bibnamefont {Pluciennik}},\ }\bibfield  {title} {\enquote {\bibinfo {title} {{Renormalization group equations for seesaw neutrino masses}},}\ }\href {\doibase 10.1016/0370-2693(93)90330-K} {\bibfield  {journal} {\bibinfo  {journal} {Phys. Lett. B}\ }\textbf {\bibinfo {volume} {316}},\ \bibinfo {pages} {312--317} (\bibinfo {year} {1993})},\ \Eprint {http://arxiv.org/abs/hep-ph/9306333} {arXiv:hep-ph/9306333} \BibitemShut {NoStop}%
\bibitem [{\citenamefont {Babu}\ \emph {et~al.}(1993)\citenamefont {Babu}, \citenamefont {Leung},\ and\ \citenamefont {Pantaleone}}]{Babu:1993qv}%
  \BibitemOpen
  \bibfield  {author} {\bibinfo {author} {\bibfnamefont {K.~S.}\ \bibnamefont {Babu}}, \bibinfo {author} {\bibfnamefont {Chung~Ngoc}\ \bibnamefont {Leung}}, \ and\ \bibinfo {author} {\bibfnamefont {James~T.}\ \bibnamefont {Pantaleone}},\ }\bibfield  {title} {\enquote {\bibinfo {title} {{Renormalization of the neutrino mass operator}},}\ }\href {\doibase 10.1016/0370-2693(93)90801-N} {\bibfield  {journal} {\bibinfo  {journal} {Phys. Lett. B}\ }\textbf {\bibinfo {volume} {319}},\ \bibinfo {pages} {191--198} (\bibinfo {year} {1993})},\ \Eprint {http://arxiv.org/abs/hep-ph/9309223} {arXiv:hep-ph/9309223} \BibitemShut {NoStop}%
\bibitem [{\citenamefont {Antusch}\ \emph {et~al.}(2005)\citenamefont {Antusch}, \citenamefont {Kersten}, \citenamefont {Lindner}, \citenamefont {Ratz},\ and\ \citenamefont {Schmidt}}]{Antusch:2005gp}%
  \BibitemOpen
  \bibfield  {author} {\bibinfo {author} {\bibfnamefont {Stefan}\ \bibnamefont {Antusch}}, \bibinfo {author} {\bibfnamefont {J\"orn}\ \bibnamefont {Kersten}}, \bibinfo {author} {\bibfnamefont {Manfred}\ \bibnamefont {Lindner}}, \bibinfo {author} {\bibfnamefont {Michael}\ \bibnamefont {Ratz}}, \ and\ \bibinfo {author} {\bibfnamefont {Michael~Andreas}\ \bibnamefont {Schmidt}},\ }\bibfield  {title} {\enquote {\bibinfo {title} {{Running neutrino mass parameters in see-saw scenarios}},}\ }\href {\doibase 10.1088/1126-6708/2005/03/024} {\bibfield  {journal} {\bibinfo  {journal} {JHEP}\ }\textbf {\bibinfo {volume} {03}},\ \bibinfo {pages} {024} (\bibinfo {year} {2005})},\ \Eprint {http://arxiv.org/abs/hep-ph/0501272} {arXiv:hep-ph/0501272} \BibitemShut {NoStop}%
\bibitem [{\citenamefont {Mei}(2005)}]{Mei:2005qp}%
  \BibitemOpen
  \bibfield  {author} {\bibinfo {author} {\bibfnamefont {Jian-wei}\ \bibnamefont {Mei}},\ }\bibfield  {title} {\enquote {\bibinfo {title} {{Running neutrino masses, leptonic mixing angles and CP-violating phases: From M(Z) to Lambda(GUT)}},}\ }\href {\doibase 10.1103/PhysRevD.71.073012} {\bibfield  {journal} {\bibinfo  {journal} {Phys. Rev. D}\ }\textbf {\bibinfo {volume} {71}},\ \bibinfo {pages} {073012} (\bibinfo {year} {2005})},\ \Eprint {http://arxiv.org/abs/hep-ph/0502015} {arXiv:hep-ph/0502015} \BibitemShut {NoStop}%
\bibitem [{\citenamefont {Dorsner}(2012)}]{Dorsner:2012uz}%
  \BibitemOpen
  \bibfield  {author} {\bibinfo {author} {\bibfnamefont {Ilja}\ \bibnamefont {Dorsner}},\ }\bibfield  {title} {\enquote {\bibinfo {title} {{A scalar leptoquark in SU(5)}},}\ }\href {\doibase 10.1103/PhysRevD.86.055009} {\bibfield  {journal} {\bibinfo  {journal} {Phys. Rev. D}\ }\textbf {\bibinfo {volume} {86}},\ \bibinfo {pages} {055009} (\bibinfo {year} {2012})},\ \Eprint {http://arxiv.org/abs/1206.5998} {arXiv:1206.5998 [hep-ph]} \BibitemShut {NoStop}%
\bibitem [{\citenamefont {Cox}\ \emph {et~al.}(2017)\citenamefont {Cox}, \citenamefont {Kusenko}, \citenamefont {Sumensari},\ and\ \citenamefont {Yanagida}}]{Cox:2016epl}%
  \BibitemOpen
  \bibfield  {author} {\bibinfo {author} {\bibfnamefont {Peter}\ \bibnamefont {Cox}}, \bibinfo {author} {\bibfnamefont {Alexander}\ \bibnamefont {Kusenko}}, \bibinfo {author} {\bibfnamefont {Olcyr}\ \bibnamefont {Sumensari}}, \ and\ \bibinfo {author} {\bibfnamefont {Tsutomu~T.}\ \bibnamefont {Yanagida}},\ }\bibfield  {title} {\enquote {\bibinfo {title} {{SU(5) Unification with TeV-scale Leptoquarks}},}\ }\href {\doibase 10.1007/JHEP03(2017)035} {\bibfield  {journal} {\bibinfo  {journal} {JHEP}\ }\textbf {\bibinfo {volume} {03}},\ \bibinfo {pages} {035} (\bibinfo {year} {2017})},\ \Eprint {http://arxiv.org/abs/1612.03923} {arXiv:1612.03923 [hep-ph]} \BibitemShut {NoStop}%
\end{thebibliography}%

\end{document}